\documentclass[aps,prb,twocolumn,showpacs]{revtex4-1}
\usepackage{graphicx}
\usepackage{amsmath,amssymb,amsfonts}
\usepackage[breaklinks]{hyperref}
\usepackage{bm}

\newcommand{\Ham}{\mathcal{H}}
\newcommand{\beq}{\begin{eqnarray}}
\newcommand{\eeq}{\end{eqnarray}}

\begin{document}

\title{Orbitally and Magnetically Induced Anisotropy in Iron-based Superconductors}
\author{Weicheng Lv}
\affiliation{Department of Physics, University of Illinois, 1110 West Green Street, Urbana, Illinois 61801, USA}
\author{Philip Phillips}
\affiliation{Department of Physics, University of Illinois, 1110 West Green Street, Urbana, Illinois 61801, USA}
\date{\today}

\begin{abstract}
Recent experimental developments in the iron pnictides have
unambiguously demonstrated the existence of in-plane electronic
anisotropy in the absence of the long-range magnetic order. Such anisotropy can
arise from orbital ordering, which is described by an energy
splitting between the two otherwise degenerate $d_{xz}$ and
$d_{yz}$ orbitals. By including this phenomenological orbital order into a five-orbital Hubbard model, we obtain the mean-field solutions where the magnetic
order is determined self-consistently.
Despite sensitivity of the resulting states to the input parameters,
we find that a weak orbital order that places the $d_{yz}$
orbital slightly higher in energy than the $d_{xz}$ orbital, combined
with intermediate on-site interactions, produces band dispersions that
are compatible with the photoemission results. In this
regime, the stripe antiferromagnetic order is further stabilized and
the resistivity displays the observed anisotropy.  We
also calculate the optical conductivity and show that it
agrees with the temperature evolution of the anisotropy seen experimentally.
\end{abstract}


\maketitle

\section{Introduction}
\label{sec:intro}

In-plane anisotropy plays a prominent role in
iron-based superconductors. For example, inelastic neutron scattering (INS)
first established that the exchange constant along the ferromagnetic
$y$ direction, $J_{1y}$, is not only much smaller than the one in the
antiferromagnetic $x$ direction, $J_{1x}$, but is actually slightly
negative,\cite{Zhao2009} suggesting that an unfrustrated spin model
underlies the magnetism.
Angle-resolved photoemission spectroscopy (ARPES) also observed a dramatic orbital-dependent Fermi-surface reconstruction upon the magnetostructural phase transition.\cite{Shimojima2010,Wang2010}
However, due to the fact that the crystals used in such experiments spontaneously form dense domains, the signals from the two diagonal phases were mixed in these  early experiments.
Hence, it was crucial that scanning tunneling microscopy (STM) detected a quasi-one-dimensional interference pattern,\cite{Chuang2010} thus confirming that the anisotropy arises entirely from a single domain.
Helping to complete the story were the breakthrough experiments in which a uniaxial stress was applied to almost fully detwin the crystals.\cite{Chu2010} This technique revealed an in-plane anisotropy in both the dc \cite{Chu2010,Tanatar2010,Kuo2011,Ying2011} and ac \cite{Dusza2011,Nakajima2011,Lucarelli2011} conductivities.
Consistent with this transport anisotropy are the band dispersions derived from ARPES measurements on detwinned samples.\cite{Yi2011}

The observed anisotropies (for a review, see Ref.~\onlinecite{Fisher2011}) have distinct origins in terms of the
local-itinerant electron dichotomy.\cite{Wu2008,Dai2009,Kou2009}
On the one hand, INS measures the spin excitations that arise predominantly
from local moments. Although a structural transition precedes or coincides with the onset of the antiferromagnetic order, its magnitude is too small to account for the strong anisotropy.
In this context, several theories
\cite{Kruger2009,Lv2009,Lee2009,Chen2009} adopted a Kugel-Khomskii \cite{Kugel1982}
model with orbital-dependent superexchanges and proposed that
ferro-orbital order stabilizes the $(\pi,0)$ magnetic order, leading naturally to anisotropic spin couplings.
In an alternative approach,\cite{Lv2010,Yin2010} the local moments, which are
governed by an isotropic $J_1$-$J_2$ Heisenberg model, are coupled
with the itinerant electrons of the two degenerate $d_{xz}$ and $d_{yz}$ orbitals by the Hund exchange. From the double-exchange mechanism, a ferro-orbital order in the itinerant band leads to different corrections to the spin-wave excitations along two diagonal directions, resulting in an anisotropic spin-only model.

On the other hand, the detected anisotropy could simply signify that the itinerant electrons also undergo a phase transition that breaks the $C_4$ rotational symmetry, which is no surprise since the symmetry is already broken by the underlying $(\pi,0)$ antiferromagnetism.
Although this electronic anisotropy does persist above the N\'eel temperature $T_N$ and the structural transition temperature $T_S$ in some experiments,\cite{Chu2010,Tanatar2010,Kuo2011,Ying2011,Dusza2011,Nakajima2011,Lucarelli2011,Yi2011} it should be noted that the applied uniaxial stress unfortunately turns the sharp transition into a broad crossover.
In this sense, recent INS measurements \cite{Harriger2011} that
observed a spin nematic order in the paramagnetic phase have taught
us something new -- the magnetic anisotropy, and the possible
electronic anisotropy, can exist in the absence of the $(\pi,0)$
antiferromagnetism and even the structural distortion. This
``nematic'' order is indeed confirmed by the band splitting above $T_S$
observed by the latest ARPES experiments on twinned samples,\cite{Yi2011} and recently hinted at by the zero-bias conductance enhancement in point-contact spectroscopy.\cite{Arham2011} More convincing evidence comes from the anisotropic vortex core structure in the superconducting FeSe,\cite{Song2011} which exhibits no static spin order.

One theoretical proposal that is in accord with some of the current experimental
observations is the $J_1$-$J_2$ Heisenberg model,\cite{Si2008} in
which an Ising order can occur without long-range magnetic
order,\cite{Xu2008,Fang2008} according to the ``order from disorder without order''
mechanism.\cite{Chandra1990a} However, the unfrustrated magnetism
observed by INS\cite{Zhao2009,Harriger2011} suggests that the correct
physical model lies elsewhere.
Another possible candidate is a Kugel-Khomskii spin-orbital model \cite{Kruger2009} that might support a phase which possesses some orbital order that breaks the $C_4$ rotational symmetry, but no long-range magnetic order. Besides these strong-coupling approaches, there are other theoretical proposals that attribute the nematic order to interaction-induced instability of the itinerant electrons.\cite{Zhai2009,Kang2011,Fernandes2011a}

Our particular view, which we have enunciated
elsewhere,\cite{Lv2009,Lv2010} is that orbital order (rather than any
intrinsic tendency of the electrons to orient themselves
inhomogeneously) lies at the heart of the
anisotropies and the observed structural and magnetic
transitions. Certainly, it is now common\cite{Kivelson1998} to refer to any phase that breaks
$C_4$ symmetry as an electron nematic, even when such a phase is
lattice or orbitally
induced. This view has been scrutinized sharply through recent
experiments\cite{Zeljkovic2011} on the the cuprate superconductor Bi$_2$Sr$_2$Ca$_{n-1}$Cu$_n$O$_{2n+4+x}$, which indicate that many of
the anisotropies that have been viewed as evidence for an electron
nematic phase actually originate from lattice effects in the BiO
layer. This experiment certainly indicates that caution rather than a
rush to nematize is warranted.  For the pnictides, the origin of the anisotropies
will remain open in the absence of a clear experiment that is able to
discern their efficient cause. Consequently, our
usage of the term nematic here strictly entails the orbital
order which necessarily breaks the equivalence between the canonical
$x$ and $y$ axes.

In this paper, we will not focus on the origin of this ``nematic'' order
(in the sense defined above), but rather investigate its experimental consequences by modeling it phenomenologically as an energy-splitting term between the Fe $d_{xz}$ and $d_{yz}$ orbitals.
This type of orbital nematic order has been studied
previously,\cite{Chen2010} but only in the context of an
orbital-independent magnetic order, which is insufficient to capture
the complicated electronic structure. To this end, we will start with
the multi-orbital model that explicitly includes this orbital nematic
order and solve the mean-field Hamiltonian using a self-consistent
procedure. In this approach, the magnetic order on different orbitals
will be determined more realistically by the band structure and the
interaction strength. Moreover, we will be able to address how the
orbital and magnetic orders interplay with one another. To reach agreement with photoemission experiments,\cite{Yi2011} we find that the $d_{xz}$ orbital is placed lower in energy than $d_{yz}$, and that this orbital order strengthens the stripe antiferromagnetism. The
orbital and magnetic order together reconstruct the band structure and
result in the anisotropy at both the low-temperature antiferromagnetic
and the high-temperature paramagnetic phase. These findings suggest
that orbital order plays the central role in the electronic structure of the iron-based superconductors.

The paper is organized as follows. The general formalism of the
multi-orbital model is described in Sec.~\ref{sec:model}. We introduce
the orbital nematic order in Sec.~\ref{sec:orbital} and calculate its
experimental consequences that are relevant for anisotropy in the
paramagnetic phase. Section \ref{sec:magnetic} presents the complicated
electronic structure arising from both the orbital and magnetic
orders. It is shown that both of them are indispensable components
that lead to anisotropies observed by many experiments in the
magnetically ordered state. The implications of our findings
are discussed in the last section.

\section{Multi-orbital model}
\label{sec:model}

The multi-orbital Hamiltonian we start with is usually defined within
an extended Brillouin zone that only contains one Fe atom per unit
cell. The kinetic-energy contribution is written as
\beq
    \Ham_K = \sum_{\alpha\beta} \sum_{\tilde{\bm{k}}\mu} \hat{\xi}_{\alpha\beta} (\tilde{\bm{k}}) c^\dagger_{\tilde{\bm{k}}\alpha\mu} c_{\tilde{\bm{k}}\beta\mu},
\label{eq:HK}
\eeq
where $c^\dagger_{\tilde{\bm{k}}\alpha\mu}$ creates an electron of momentum $\tilde{\bm{k}}$ on orbital $\alpha$ with spin $\mu$ ($\mu=\uparrow,\downarrow$).
We note here that $\tilde{\bm{k}}$ is not the crystal momentum $\bm{k}$
defined by the translation operator $\mathcal{T}_i$ of the Fe lattice
unit vector $e_i$ ($i = x,y,z$), i.e., $\mathcal{T}_i \vert \bm{k} \rangle =
e^{i k_i} \vert \bm{k} \rangle$. Rather, it is the operator $\mathcal{P}_z \mathcal{T}_i$, where $\mathcal{P}_z$ is the reflection operator in the $z$ direction, instead of $\mathcal{T}_i$, that leaves the Fe-As lattice invariant.
Thus, $\tilde{\bm{k}}$ actually labels the eigenstates of
$\mathcal{P}_z \mathcal{T}_i$, i.e., $\mathcal{P}_z \mathcal{T}_i \vert
\tilde{\bm{k}} \rangle = e^{i \tilde{k}_i} \vert \tilde{\bm{k}}
\rangle$, and yields an unambiguous way to unfold the real Brillouin
zone with two Fe atoms per unit cell. This important distinction
between $\tilde{\bm{k}}$ and $\bm{k}$ was previously discussed in
detail in Ref.~\onlinecite{Lee2008}.
In principle, we need to fold back the band dispersions that are
obtained by the diagonalization of $\Ham_K$ (\ref{eq:HK}), and fit them
to the local-density-approximation (LDA) calculations by tuning the
tight-binding hopping parameters. This has been done in many studies, using two,\cite{Raghu2008,Ran2009,Moreo2009} three,\cite{Lee2008,Daghofer2010} four,\cite{Yu2009} or five\cite{Kuroki2008,Graser2009,Calderon2009,Eschrig2009,Graser2010,Luo2010} Fe $d$ orbitals. In order to make our calculations more realistic, we only focus on the five-orbital model, particularly the one of Ref.~\onlinecite{Graser2010}, which is based on a three-dimensional fitting to the LDA band structures of BaFe$_2$As$_2$, the material on which most of the experiments are performed.

It is helpful for us to return to real space where $\Ham_K$ (\ref{eq:HK}) takes the form $\Ham_K = \sum t_{ij}^{\alpha\beta} c_{i\alpha\mu}^\dagger c_{j\beta\mu}$,
where $t_{ij}^{\alpha\beta}$ is the hopping amplitude, with $i$, $j$ denoting the index of the site. As discussed, the operator $\mathcal{P}_z \mathcal{T}_i$ leaves $\Ham_K$ invariant. Under $\mathcal{P}_z \mathcal{T}_i$, we have $c_{i\alpha\mu} \rightarrow \chi_\alpha c_{i+e_i,\alpha\mu}$, where $\chi_\alpha=1$ for $\alpha = d_{xy}$, $d_{x^2-y^2}$, or $d_{3z^2-r^2}$, and $\chi_\alpha = -1$ for $\alpha = d_{xz}$ or $d_{yz}$.
Thus, it is required that $t_{ij}^{\alpha\beta} = t^{\alpha\beta}_{i-j}$ for $\chi_\alpha \chi_\beta = 1$, whereas $t_{ij}^{\alpha\beta} = e^{i\bm{K}\cdot\bm{r}_i} t^{\alpha\beta}_{i-j}$ for $\chi_\alpha \chi_\beta = -1$, where $\bm{K} = (\pi,\pi,\pi)$ and $t^{\alpha\beta}_{i-j}$ only depends on $\bm{r}_i-\bm{r}_j$.
We immediately noticed that in the crystal momentum space, the
electron operators of the $d_{xz}$ and $d_{yz}$ orbitals at $\bm{k}$
are coupled with those of $d_{xy}$, $d_{x^2-y^2}$, and $d_{3z^2-r^2}$ at
$\bm{k}+\bm{K}$. As a result, we define the pseudocrystal momentum
$\tilde{\bm{k}}$ as follows: $c_{\tilde{\bm{k}}\alpha\mu} =
c_{\bm{k}\alpha\mu}$ for $\alpha = d_{xz}$ or $d_{yz}$, and
$c_{\tilde{\bm{k}}\alpha\mu} = c_{\bm{k}+\bm{K},\alpha\mu}$ for
$\alpha = d_{xy}$, $d_{x^2-y^2}$, or
$d_{3z^2-r^2}$.\cite{Lee2008,Yu2009,Calderon2009}  It is in this pseudocrystal momentum $\tilde{\bm{k}}$ space that the kinetic energy $\Ham_K$ takes the diagonal form of $\Ham_K$ (\ref{eq:HK}).

The distinction between $\tilde{\bm{k}}$ and $\bm{k}$ has immediate consequences for the interpretation of the ARPES measurements. The momentum probed by ARPES is not $\tilde{\bm{k}}$, but $\bm{k}$. Because the crystal momentum $\bm{k}$ is not a good quantum number of $\Ham_K$~(\ref{eq:HK}), ARPES detects both bands with momentum $\tilde{\bm{k}}$ and $\tilde{\bm{k}}+\bm{K}$, corresponding to folding the Brillouin zone with one Fe atom per unit cell by a wave vector $\bm{K} = (\pi,\pi,\pi)$. This observation is consistent with the fact that the real unit cell including the As atoms consists of two Fe atoms. However, the relative intensity of each band measured by ARPES depends on the strengths of hybridizations between $\bm{k}$ and $\bm{k}+\bm{K}$.

Formally, the Green function in terms of $\tilde{\bm{k}}$ is defined as
\beq
    \hat{\mathcal{G}}_{\alpha\beta} (\tilde{\bm{k}},\tau) = - \left\langle T_\tau c_{\tilde{\bm{k}}\alpha\mu}(\tau) c_{\tilde{\bm{k}}\beta\mu}^\dagger(0) \right\rangle,
\eeq
with its Fourier transform satisfying $\hat{\mathcal{G}}^{-1} (\tilde{\bm{k}},\omega) = \omega \hat{I} - \hat{\xi}(\tilde{\bm{k}})$.
We can simply write down the spectral function
\beq
A(\tilde{\bm{k}},\omega) = - \frac{1}{\pi} \Im \left[ \mathrm{tr} \, \hat{\mathcal{G}} (\tilde{\bm{k}},\omega+ i\delta ) \right].
\eeq
However, what is really measured by ARPES is not $A(\tilde{\bm{k}},\omega)$, but
\beq
A(\bm{k},\omega)  & = & - \frac{1}{\pi} \Im \left[ \sum_{\chi_\alpha=-1} \hat{\mathcal{G}}_{\alpha\alpha} (\tilde{\bm{k}},\omega+ i\delta ) \right. \nonumber \\
& & + \, \left. \sum_{\chi_\alpha=1} \hat{\mathcal{G}}_{\alpha\alpha} (\tilde{\bm{k}}+\bm{K},\omega+ i\delta ) \right].
\eeq
It is this function on which we will focus.

\begin{figure}
  \centering
  \includegraphics[width=8cm]{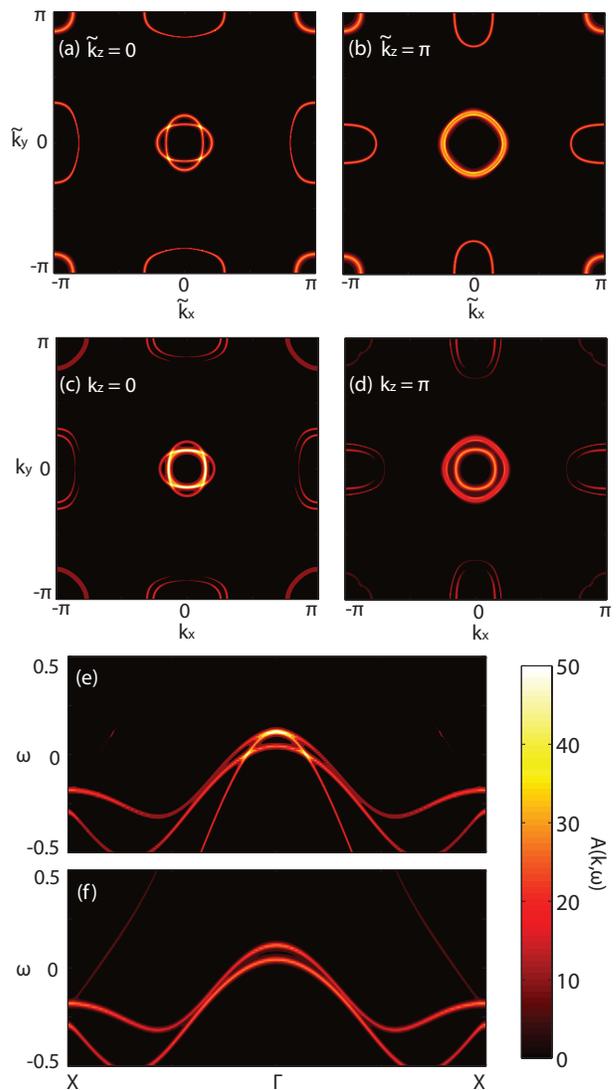}
  \caption{(Color online) Fermi surfaces in terms of (a), (b) the
    pseudocrystal momentum $\tilde{\bm{k}}$ and (c), (d) the crystal
    momentum $\bm{k}$. We plot the Fermi surfaces in the $xy$ plane
    and set the $z$ component in each figure to (a) $\tilde{k}_z = 0$,
    (b) $\tilde{k}_z = \pi$, (c) $k_z = 0$, and (d) $k_z = \pi$. (e)
    Unpolarized dispersions and (f) polarized dispersions along the
    crystal momentum line of $X$-$\Gamma$-$X$. In (f), only the
    components of the $d_{yz}$ and $d_{xy}$ orbitals are shown. We
    represent the values of the spectral function $A(k,\omega)$ by the
    color scale, which is used consistently for all of the figures in this paper.}
  \label{fig:original}
\end{figure}

In Figs.~\ref{fig:original}(a) and (b), we plot
$A(\tilde{\bm{k}},\omega=0)$, depicting the Fermi surfaces in the
plane of $\tilde{k}_z =0$ and $\tilde{k}_z=\pi$, respectively. Those
of $A(\bm{k},\omega=0)$ are shown in Figs.~\ref{fig:original}(c) and
(d). The frequency $\omega$ is defined with respect to the chemical
potential $\mu$, which is determined by the electron filling level
$n=6$, dictated by the Fe valence in the parent compounds. In
addition, all the energy scales have an implicit unit of $eV$ throughout this paper. By comparing these graphs, we find that $A(\bm{k},\omega)$
displays a more complicated structure due to the folding of the Brillouin
zone. From now on, we will only focus on $A(\bm{k},\omega)$, which is
measured by ARPES in our direct comparison with
experiments. Furthermore, in Fig.~\ref{fig:original}(e),
$A(\bm{k},\omega)$ is plotted along the line $X$-$\Gamma$-$X$, with
$X=(\pi,0,0)$ and $\Gamma = (0,0,0)$, to represent the band
dispersions probed by unpolarized ARPES. Experimentally, the orbital
character can be investigated by tuning the polarization of the
incoming light. For example, using a polarization perpendicular to the
incident plane only selects those orbitals that are odd under
$\mathcal{P}_y$ ($d_{yz}$ and $d_{xy}$) along the line of
$X$-$\Gamma$-$X$. We plot these orbital-polarized dispersions in
Fig.~\ref{fig:original}(f), which shows qualitative agreement with
experiments.\cite{Yi2011} Since the $C_4$ rotational symmetry is
respected, the dispersions are exactly the same along the crystal momentum line $Y$-$\Gamma$-$Y$, where $Y=(0,\pi,0)$, with no splitting between the bands at $X$ and $Y$.

\section{Orbital nematic order}
\label{sec:orbital}

\begin{figure}
  \centering
  \includegraphics[width=8cm]{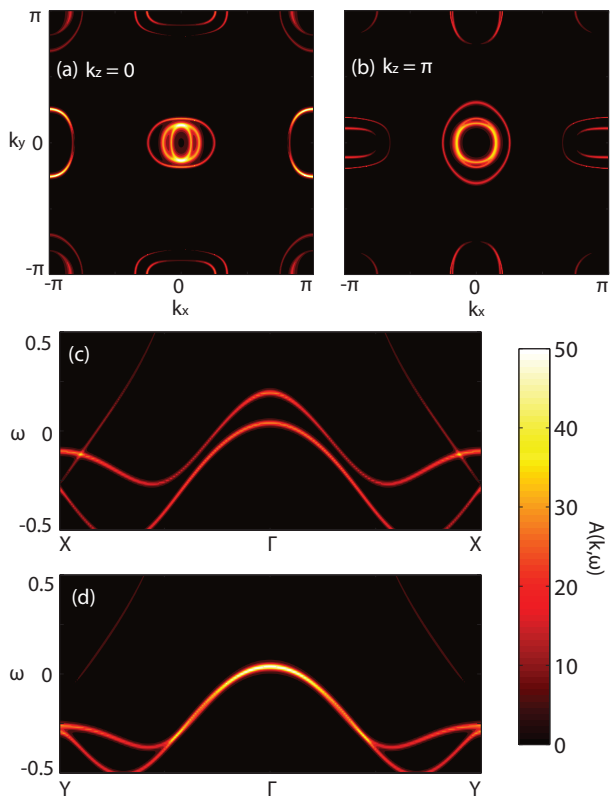}
  \caption{(Color online) Fermi surfaces in the plane for (a) $k_z = 0$ and (b) $k_z = \pi$. Polarized dispersions along the line of (c) $X$-$\Gamma$-$X$ and (d) $Y$-$\Gamma$-$Y$. We have set the orbital nematic order parameter $\Delta = -0.08$, explicitly breaking $C_4$ symmetry.}
  \label{fig:orbital}
\end{figure}

As discussed in Sec.~\ref{sec:intro}, electronic anisotropy has
been confirmed by recent experiments \cite{Yi2011,Harriger2011} to
persist above the magnetostructural transition.  Our take
on this is that this effect is due entirely to orbital ordering.  To
test out this hypothesis,  we introduce the orbital nematic order as an energy-splitting term between the $d_{xz}$ and $d_{yz}$ orbitals,
\beq
    \Ham_N = \sum_{i\alpha\mu} \Delta_\alpha  c_{i\alpha\mu}^\dagger c_{i\alpha\mu},
\label{eq:HN}
\eeq
where $\Delta_\alpha = \pm \Delta$ for $\alpha = d_{xz}$ and $d_{yz}$, respectively, and $\Delta_\alpha=0$ for the other three orbitals. In principle, all five orbitals should be involved in this nematic order. But we will only consider the $d_{xz}$ and $d_{yz}$ orbitals due to their spatial anisotropy, whereas the other three orbitals are dropped because they are $C_4$ rotationally symmetric. It needs to be emphasized that $\Ham_N$ (\ref{eq:HN}) represents an electron nematic order which occurs without the onset of the long-range stripe antiferromagnetism.
We find that in order to produce results that are consistent with
ARPES measurements,\cite{Yi2011} the orbital nematic order parameter
$\Delta$ is required to have a small negative value, which leads to a
higher energy of $d_{yz}$ relative to $d_{xz}$. For the purpose of illustration, we choose $\Delta = -0.08$, and plot the resulting Fermi surfaces and polarized dispersions in Fig.~\ref{fig:orbital}.

From Figs.~\ref{fig:orbital}(a) and (b), we notice that the hole
pockets at the zone center are deformed into elliptical shapes that
are elongated along the $x$ and $y$ direction in the plane of $k_z=0$
and $k_z=\pi$, respectively. The apparent breaking of $C_4$
symmetry is indeed a direct consequence of $\Ham_N$~(\ref{eq:HN})
that is explicitly introduced by hand. However, this orbital order does have a physically reasonable origin, which can be explained as follows.
From a weak-coupling point of view, the observed magnetic order with wave vector $\bm{q}=(\pi,0,\pi)$ arises from the nesting instability between the hole pockets centered at $\tilde{\bm{k}}=(0,0,\tilde{k}_z)$ and the electron pockets at $\tilde{\bm{k}}=(\pi,0,\tilde{k}_z+\pi)$. But the nesting is not perfect because the hole pockets are more circular, whereas the electron pockets are more elliptical [see Figs.~\ref{fig:original}(a) and (b)]. By the inclusion of $\Ham_N$ (\ref{eq:HN}) with a small negative $\Delta$, the hole pockets are deformed into ellipses, whereas the electron pockets are relatively less affected, thus resulting in a better nesting condition between the two by the wave vector $\bm{q}$ [see Figs.~\ref{fig:orbital}(a) and (b)]. Consequently, this type of orbital order will naturally arise in the system and provide further stabilization of the antiferromagnetism. Our result is consistent with the Pomeranchuk instability from the functional renormalization-group studies.\cite{Zhai2009}

The polarized dispersions along the $x$ and $y$ direction are displayed in Figs.~\ref{fig:orbital}(c) and (d), respectively. It needs to be emphasized that only the spectral functions of $d_{yz}$ and $d_{xy}$ orbitals are shown in Fig.~\ref{fig:orbital}(c), whereas we only plot those of $d_{xz}$ and $d_{xy}$ in Fig.~\ref{fig:orbital}(d), which are exactly what are measured by the polarized ARPES setup.\cite{Yi2011} Indeed, a small negative $\Delta$, which lifts $d_{yz}$ higher than $d_{xz}$ in energy, produces splitting between the bands at $X$ and $Y$, in agreement with experimental observations.

\begin{figure}
  \centering
  \includegraphics[width=8cm]{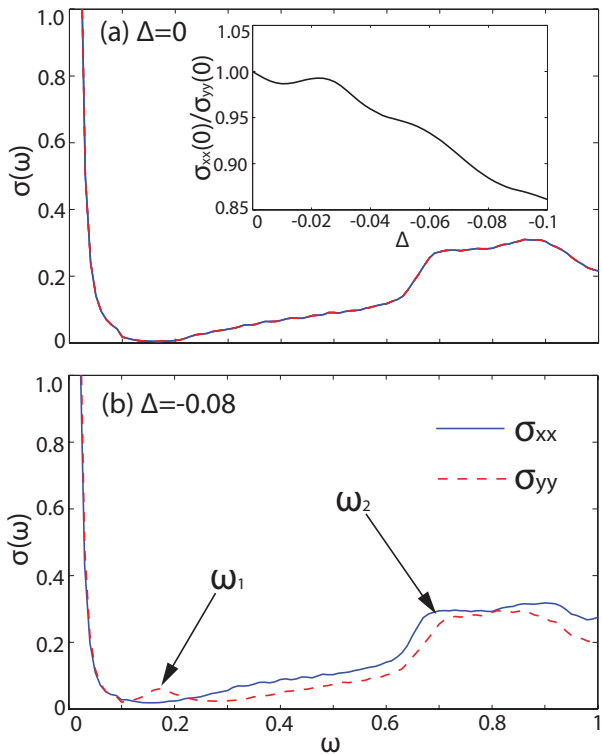}
  \caption{(Color online) Optical conductivity $\sigma_{xx}$ and
    $\sigma_{yy}$ as a function of the frequency $\omega$ for the
    orbital nematic order parameter (a) $\Delta=0$ and (b)
    $\Delta=-0.08$. The inset of (a) displays the ratio of the Drude
    weight, $\sigma_{xx}(0)/\sigma_{yy}(0)$, as a function of
    $\Delta$. $\sigma(\omega)$ is plotted in an arbitrary unit, which is kept the same in this paper. The
    two arrows in (b) denote the two characteristic frequencies
    $\omega_1 \approx 0.2$ and $\omega_2 \approx 0.7$, where $\sigma(\omega)$ exhibits a peak structure.}
  \label{fig:optical_orb}
\end{figure}

To further examine the validity of the orbital order term
$\Ham_N$ (\ref{eq:HN}), we turn to its experimental consequences in
transport measurements by calculating the optical conductivity
$\sigma_{xx}(\omega)$ and $\sigma_{yy}(\omega)$ along the $x$ and
$y$ direction, respectively. When $\Delta=0$, $\sigma_{xx}$ and
$\sigma_{yy}$ are completely equivalent
[Fig.~\ref{fig:optical_orb}(a)], preserving the $C_4$ symmetry. We see
a sharp Drude peak accompanied by some high-energy features due to
interband transitions. If a small negative $\Delta=-0.08$ is turned
on, $\sigma(\omega)$ along two diagonal directions become distinct and
a low-energy subpeak emerges around $\omega_1 \approx 0.2$, where
$\sigma_{yy}$ has a value larger than $\sigma_{xx}$, as depicted in
Fig.~\ref{fig:optical_orb}(b). We also notice that around the
high-energy peak at $\omega_2 \approx 0.7$, $\sigma_{xx}$ is dominant
instead. These results are in good agreement with experiments
\cite{Dusza2011} performed above $T_N$ in the detwinned samples.

We also compare the Drude weight of $\sigma_{xx}$ and $\sigma_{yy}$
and plot their ratio as a function of $\Delta$. As shown in the inset
of Fig.~\ref{fig:optical_orb}(a), when the energy splitting $\Delta$
gets larger, $\sigma_{xx}(0)/\sigma_{yy}(0)$ almost monotonically
decreases, resulting in a larger Drude weight along the $y$ direction. If
we naively assume proportionality between the Drude weight and the dc
conductivity, this result contradicts experimental findings in which
the antiferromagnetic $x$ direction always has a higher
conductivity\cite{Chu2010} than the ferromagnetic $y$ direction. However, as pointed
out by optical measurements,\cite{Dusza2011} the Drude weight does obtain a higher
value along the $y$ direction, and it is the scattering rate difference that dictates the higher conductivity along the $x$ direction. In this regard, to obtain the correct anisotropy of the dc
conductivity, we need to take into account some anisotropic scattering
mechanism, for example, a short-range $(\pi,0,\pi)$ magnetic order
which is supported by a recent INS experiment.\cite{Harriger2011} Indeed, this correct anisotropy of the scattering rate is obtained by a recent theory\cite{Fernandes2011} that considers scattering by anisotropic spin fluctuations in the paramagnetic phase. It
should also be mentioned that our result of the Drude weight agrees
with earlier studies\cite{Chen2010} based on the same form of the orbital order,
but using a different five-orbital tight-binding model.\cite{Graser2009}

In conclusion, our study shows that orbital order plays an
important role in modifying the electronic structure as well as the ratio of the Drude weights,
$\sigma_{xx}(0)/\sigma_{yy}(0)$, above the
onset temperature for long-range magnetic order.  In this way, we have separated the
physics that results from magnetic as opposed to orbital order.

\section{Magnetic order}
\label{sec:magnetic}

Now we set out to include magnetic order. At the mean-field level,
magnetic order can be described by
\beq
\Ham_{AF}  =  \sum_{\alpha\beta} M_{\alpha\beta} \sum_{i}  e^{i\bm{q}\cdot\bm{r}_i} \sigma^z_{\mu\nu} c_{i\alpha\mu}^\dagger c_{i\beta\nu}
\eeq
for a multi-orbital system, where we have chosen the ordering wave vector $\bm{q} = (\pi,0,\pi)$ and the spins are assumed to point along the $S^z$ direction, with $\sigma^z$ being the Pauli matrix. For a five-orbital system, the order parameters $M_{\alpha\beta}$ form a $5\times5$ Hermitian matrix, with a total of 25 independent real variables.
However, as discussed in Ref.~\onlinecite{Ran2009}, the $(\pi,0,\pi)$
magnetically ordered state is invariant under the inversion
$\mathcal{I}$ around any Fe site, reflections $\mathcal{P}_x$,
$\mathcal{P}_y$, and $\mathcal{P}_z$ along the $x$, $y$, and $z$
directions, respectively, and effective time reversal
$\mathcal{TR}^\prime$, which is a combination of time reversal and
spin reversal, $\mathcal{TR}^\prime = \mathcal{TR} \circ
\mathcal{SR}$. Under these symmetries in our five-orbital system, only
six parameters acquire nonzero real values: $M_{\alpha\alpha}$
($\alpha = d_{xz}, d_{yz}, d_{xy}, d_{x^2-y^2}, d_{3z^2-r^2}$) and
$M_{\alpha\beta} = M_{\beta\alpha}$ ($\alpha=d_{x^2-y^2}$,
$\beta=d_{3z^2-r^2}$). Hence, magnetic order obtains almost exclusively within the same orbitals, with the only exception being the orbital-off-diagonal term between the $d_{x^2-y^2}$ and $d_{3z^2-r^2}$ orbitals.

By following these discussions, we consider the on-site interaction of a multi-orbital Hubbard model,
\beq
\Ham_I & = & \frac{U}{2} \sum_{i, \alpha,\mu \neq \nu}
\hat{n}_{i \alpha \mu} \hat{n}_{i \alpha \nu} + \frac{V}{2} \sum_{i,
\alpha \neq \beta, \mu \nu} \hat{n}_{i \alpha \mu} \hat{n}_{i \beta \nu} \nonumber \\
	 & & + \, \frac{J}{2} \sum_{i, \alpha \neq \beta, \mu \nu} c_{i \alpha \mu}^\dagger c_{i \beta \nu}^\dagger c_{i \alpha \nu} c_{i \beta \mu} \nonumber \\
	 & & + \, \frac{J^\prime}{2} \sum_{i, \alpha \neq \beta, \mu \neq \nu} c_{i \alpha \mu}^\dagger c_{i \alpha \nu}^\dagger c_{i \beta \nu} c_{i \beta \mu},
\label{eq:HI}
\eeq
where $\hat{n}_{i \alpha \mu} = c_{i \alpha \mu}^\dagger c_{i \alpha \mu}$. We still assume $U=V+2J$ and $J=J^\prime$, which is not necessarily valid as the orbitals used here only share the same symmetry but do not have exactly the identical form of the atomic $d$ orbitals. We make use of the standard mean-field decoupling,
\beq
\left \langle c_{i\alpha\mu}^\dagger c_{i\beta\nu} \right \rangle =
\frac{1}{2} \left( n_\alpha + \mu m_\alpha e^{i\bm{q}\cdot \bm{r}_i} \right) \delta_{\alpha\beta} \delta_{\mu\nu},
\label{eq:decouple}
\eeq
where $\mu = \pm 1$ for up and down spins, respectively. As shown by LDA calculations,\cite{Graser2010} the Fermi surfaces are mostly composed of the $t_{2g}$ orbitals ($d_{xz}$, $d_{yz}$, and $d_{xy}$). Thus we can safely ignore the orbital-off-diagonal magnetic order between $d_{x^2-y^2}$ and $d_{3z^2-r^2}$, and use the above orbital-diagonal decoupling which captures five of the total of six nonzero mean-field antiferromagnetic order parameters.

By contrast, we point out that in Ref.~\onlinecite{Kuroki2008}, the orbitals are defined along the axes of the original unit cell, $X$ and $Y$, which are rotated by 45$^\circ$ from the $x$ and $y$ axes of the Fe lattice. Applying mean-field decoupling of Eq.~(\ref{eq:decouple}) on this model will only take account of four order parameters. Symmetry considerations impose that the orbital-diagonal elements $M_{\alpha\alpha}$ of the $d_{XZ}$ and $d_{YZ}$ orbitals are equal to each other. However, the off-diagonal element $M_{\alpha\beta} = M_{\beta\alpha}$ ($\alpha=d_{XZ}$, $\beta=d_{YZ}$) can acquire nonzero values, but will not be captured by the mean-field theory. Hence, we will use the model\cite{Graser2010} where the orbitals are defined along the Fe-Fe bond, and expect better results compared to earlier studies\cite{Kaneshita2009,Sugimoto2011} based on Ref.~\onlinecite{Kuroki2008} under orbital-diagonal mean-field decoupling.

Straightforward calculation yields the mean-field interaction term
\beq
    \Ham_I & = & C + \sum_{\bm{k}\alpha\mu} \left[ \epsilon_\alpha
c_{\bm{k}\alpha\mu}^\dagger c_{\bm{k}\alpha\mu} \right. \nonumber \\
& & + \, \left. \eta_{\alpha\mu} \left( c_{\bm{k}\alpha\mu}^\dagger c_{\bm{k}+\bm{q},\alpha \mu} + h.c. \right) \right],
\label{eq:HIk}
\eeq
where
\beq
    \epsilon_\alpha & = & \frac{U}{2} n_\alpha + \left(V-\frac{J}{2} \right)
\sum_{\beta \neq \alpha} n_\beta, \\
    \eta_{\alpha\mu} & = & -\frac{\mu}{2} \left( U m_\alpha + J\sum_{\beta
\neq \alpha} m_\beta \right),
\eeq
and the constant
\beq
C  & = & -\frac{U}{4} \sum_\alpha \left(n_\alpha^2-m_\alpha^2 \right) - \frac{2V -J}{4} \sum_{\alpha \neq \beta} n_\alpha n_\beta \nonumber \\
& & + \, \frac{J}{4} \sum_{\alpha \neq \beta} m_\alpha m_\beta.
\eeq
Note that in $\Ham_I$ (\ref{eq:HIk}), $\bm{k}$ can be simply replaced by
$\tilde{\bm{k}}$ without changing the form of the equation. The full
Hamiltonian $\Ham = \Ham_K + \Ham_N + \Ham_I$, is quadratic in
electron operators $c_{\tilde{\bm{k}}\alpha\mu}$, and can be solved
with order parameters $n_\alpha$ and $m_\alpha$ being determined self-consistently.

\begin{figure}
  \centering
  \includegraphics[width=8cm]{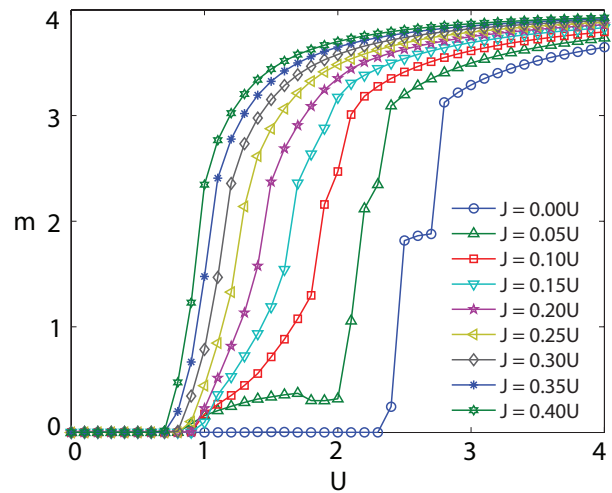}
  \caption{(Color online) The total staggered magnetic moment $m$ as a
    function of the Coulomb repulsion $U$ for different Hund's
    exchanges $J$. We set the energy splitting $\Delta =0$ in order to
    find the regime of parameters that are of interest in the context
    of iron-based superconductors.}
  \label{fig:moments}
\end{figure}

So far there is still no systematic mean-field study of this
three-dimensional tight-binding model \cite{Graser2010} specially
constructed for BaFe$_2$As$_2$. Thus, as the first step, we need to
search for appropriate values of $U$ and $J$ that are consistent with
experimental observations. At the outset, we set the orbital nematic
order $\Delta$ to zero to simplify our calculations. Generally, a
nonzero $\Delta$ produces further modifications, but the physically
relevant regime of $U$ and $J$ is not greatly affected by the choice of $\Delta$. In Fig.~\ref{fig:moments}, we plot the total staggered magnetic moment $m=\sum_{\alpha}m_\alpha$ as a function of Coulomb repulsion $U$ for various Hund's couplings $J$. It is found that there exists a metallic phase with antiferromagnetic order at intermediate Coulomb repulsion $U$. Furthermore the ratio of $J$ and $U$ also needs to take intermediate values. This requirement of $U$ and $J$ is qualitatively consistent with earlier studies \cite{Bascones2010,Luo2010} based on other five-orbital models.

\begin{figure}
  \centering
  \includegraphics[width=8cm]{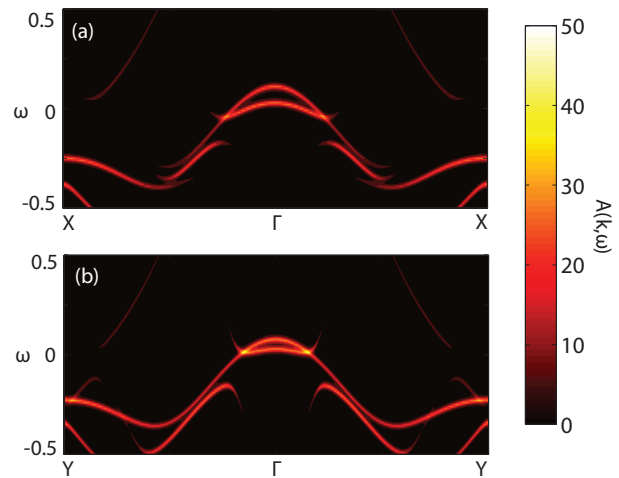}
  \caption{(Color online) Polarized dispersions along (a) $X$-$\Gamma$-$X$ and (b) $Y$-$\Gamma$-$Y$, for $U=1.08$, $J=0.20U$, and $\Delta=0$. The corresponding magnetic moment $m=0.46$.}
  \label{fig:magnetic}
\end{figure}

For a typical value of $U=1.08$ and $J=0.20U$, we plot the polarized
dispersions along the two diagonal directions in
Figs.~\ref{fig:magnetic}(a) and (b), respectively. Clearly, the
magnetic order opens up a gap close to Fermi energy and significantly
modifies the original band structure shown in
Fig.~\ref{fig:original}. However, this $(\pi,0,\pi)$
antiferromagnetism, although breaking the $C_4$ symmetry, produces
only a small splitting between the bands at $X$ and $Y$. In fact, the
band energy at $Y$ is even slightly higher than that at $X$. From our calculation, using other values of $U$ and $J$,
this near degeneracy of the bands at $X$ and $Y$ is quite robust,
which contradicts ARPES measurements.\cite{Yi2011} As will be seen,
this problem can be alleviated by including orbital ordering.
Hence, orbital order $\Ham_N$~(\ref{eq:HN}) is still present in the antiferromagnetic phase and has important experimental consequences.

\begin{figure}
  \centering
  \includegraphics[width=8cm]{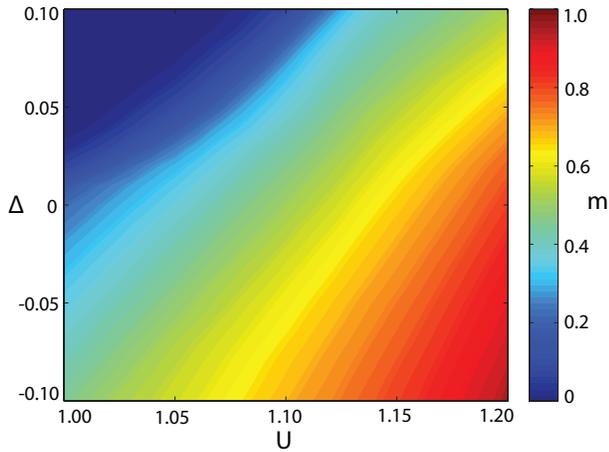}
  \caption{(Color online) The total staggered magnetic moment as a
    function of the Coulomb repulsion $U$ and orbital nematic order
    parameter $\Delta$. We set Hund's exchange $J=0.20U$.}
  \label{fig:moments_orb}
\end{figure}

In order to simplify our discussions, we set $J=0.20U$ and focus in
the regime $U \in [1.0,1.2]$ and $\Delta \in [-0.1,0.1]$, where
significant agreement with experiment can be found.   We first
investigate variations of the total staggered magnetic moment $m$ for
different $U$ and $\Delta$. As discussed in Sec.~\ref{sec:orbital}, a
negative $\Delta$ induces better nesting between hole and electron
pockets and stabilizes the $(\pi,0,\pi)$ magnetic order, thus leading
to an increase of $m$, as shown in Fig.~\ref{fig:moments_orb}. In
contrast, the magnetic moment $m$ decreases when $\Delta$ takes larger
positive values. In fact, a positive $\Delta$ places the $d_{xz}$
orbital higher in energy compared to $d_{yz}$, and favors the
antiferromagnetism with $\bm{q} = (0,\pi,\pi)$ instead. This
interesting interplay between the orbital and magnetic order was also noted recently\cite{Nevidomskyy2011} based on LDA results and a Ginzburg-Landau phenomenological theory.

\begin{figure}
  \centering
  \includegraphics[width=8cm]{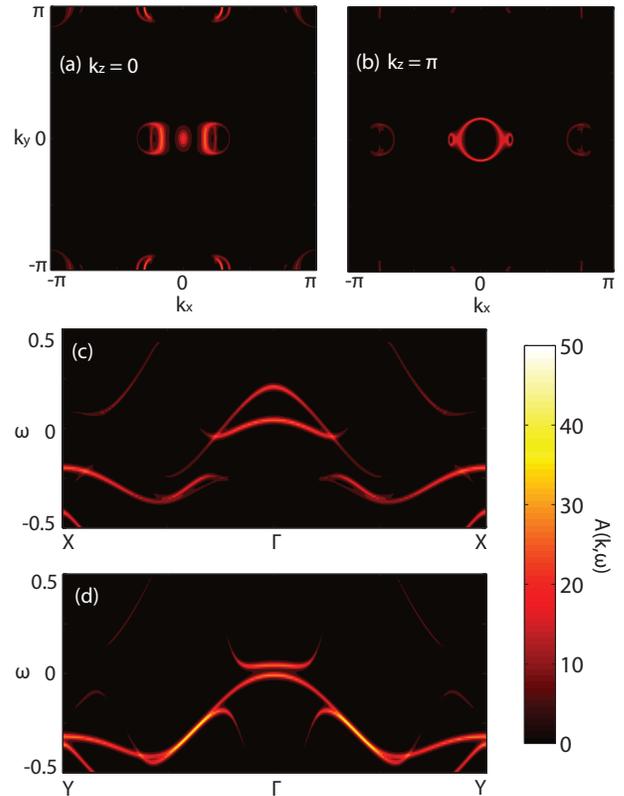}
  \caption{(Color online) Fermi surfaces in the plane of (a) $k_z = 0$ and (b) $k_z = \pi$. Polarized dispersions along the line of (c) $X$-$\Gamma$-$X$ and (d) $Y$-$\Gamma$-$Y$. The parameters used here are $U=1.08$, $J=0.20U$, and $\Delta=-0.08$, with the total mean-field staggered moment $m=0.62$.}
  \label{fig:mag_orb}
\end{figure}

To further illustrate the role of orbital order in the reconstruction
of the electronic structure in the magnetically ordered phase, we plot
the Fermi surfaces and polarized dispersions in
Fig.~\ref{fig:mag_orb}. We use the set of parameters, $U=1.08$,
$J=0.20U$, and $\Delta=-0.08$, to be consistent with previous
figures. From Figs.~\ref{fig:mag_orb}(c) and (d), the desired band
splitting between $X$ and $Y$ is successfully produced due to a
negative $\Delta$. Furthermore, there are multiple band crossings at the
Fermi energy along $X$-$\Gamma$-$X$, whereas only a single crossing
occurs along $Y$-$\Gamma$-$Y$, also in agreement with ARPES
findings.\cite{Yi2011}
We further point out that the two small Fermi
surfaces adjacent to the large hole pocket in
Fig.~\ref{fig:mag_orb}(b) are actually Dirac cones, which have been
predicted theoretically \cite{Ran2009} and confirmed
experimentally.\cite{Richard2010,Yi2011} Note, however, that the
existence of Dirac cones relies on the degeneracy between the $d_{xz}$
and $d_{yz}$ orbitals.\cite{Ran2009} But the small orbital order
used here is not enough to annihilate such Dirac features. Finally, the
two Fermi-surface segments close to the zone center in Fig.~\ref{fig:mag_orb}(a) are mostly aligned along the ferromagnetic $y$ direction and in principle can produce the quasi-one-dimensional interference pattern observed in STM.\cite{Chuang2010}

\begin{figure}
  \centering
  \includegraphics[width=8cm]{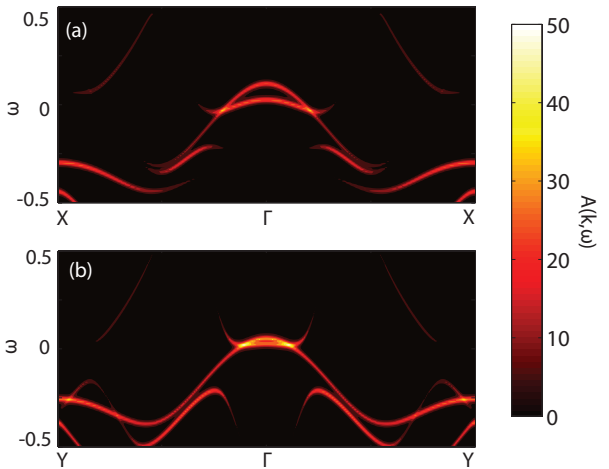}
  \caption{(Color online) Polarized dispersions along (a) $X$-$\Gamma$-$X$ and (b) $Y$-$\Gamma$-$Y$, for $U=1.13$, $J=0.20U$, and $\Delta=0$. The corresponding magnetic moment $m=0.61$, which is close to $m=0.62$ of Fig.~\ref{fig:mag_orb}.}
  \label{fig:mag_no_orb}
\end{figure}

However, attentive readers may notice that for Fig.~\ref{fig:mag_orb}, the corresponding magnetic moment $m=0.62$, which is larger than $m=0.46$ of Fig.~\ref{fig:magnetic}. Hence, it is possible that the splitting between the bands at $X$ and $Y$ is caused by the stronger magnetic order in Fig.~\ref{fig:mag_orb}. In order to confirm the splitting is entirely due to the orbital order term $\Ham_N$ (\ref{eq:HN}), we plot the polarized dispersions in Fig.~\ref{fig:mag_no_orb} for $U=1.13$, $J=0.20U$, and $\Delta=0$. These parameters produce a staggered magnetic moment $m=0.61$, which is close to the value of $m=0.62$ in Fig.~\ref{fig:mag_orb}. However, no splitting between the bands at $X$ and $Y$ is generated in Fig.~\ref{fig:mag_no_orb}, where the band energy at $X$ is even slightly lower than that at $Y$. As we have pointed out earlier, this near degeneracy between the band energy at $X$ and $Y$ survives for other values of $U$ and $J$ as well, as long as the orbital nematic order parameter $\Delta=0$. Therefore we have shown that it is the orbital order $\Ham_N$ (\ref{eq:HN}), rather than the stripe antiferromagnetism, that is responsible for the band splitting at $X$ and $Y$ observed in ARPES.\cite{Yi2011}

\begin{figure}
  \centering
  \includegraphics[width=8cm]{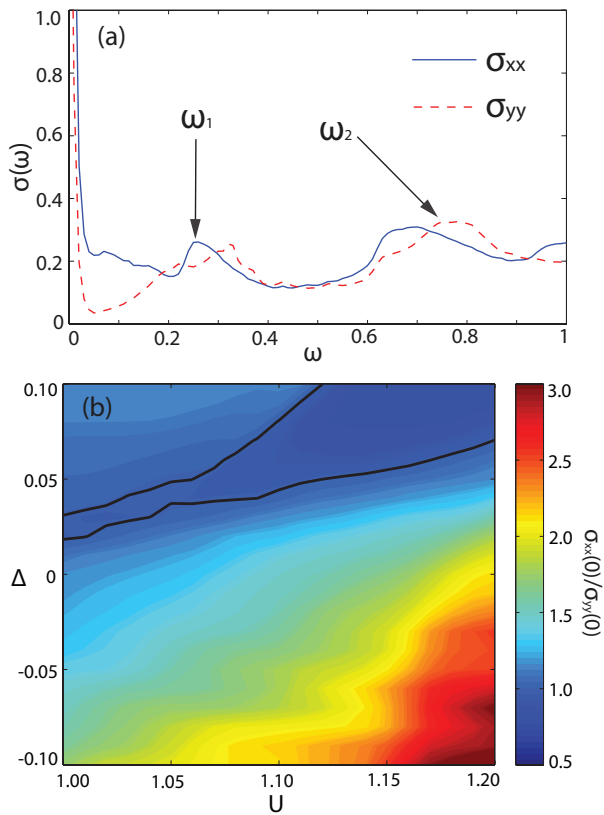}
  \caption{(Color online) (a) Optical conductivity
    $\sigma_{xx}(\omega)$ and $\sigma_{yy}(\omega)$ calculated using
    $U=1.08$, $J=0.20U$, and $\Delta=-0.08$. They are plotted in the same unit as Fig.~\ref{fig:optical_orb}. (b) Anisotropy of Drude
    weight, $\sigma_{xx}(0)/\sigma_{yy}(0)$, as a function of  $U$ and
    $\Delta$. The black line separates the regions where $\sigma_{xx}(0) > \sigma_{yy}(0)$ and where $\sigma_{xx}(0) < \sigma_{yy}(0)$. We choose $J=0.20U$ here.}
  \label{fig:optical_mag_orb}
\end{figure}

We end this section by discussing the optical spectra in the presence
of both orbital and magnetic order. The same set of parameters as
those of Fig.~\ref{fig:mag_orb} are used here. From
Fig.~\ref{fig:optical_mag_orb}(a), we see that the optical
conductivity shows two peaks at frequency $\omega_1$ and $\omega_2$,
which have similar values as we defined previously. Compared to the
results without the magnetic order (see Fig.~\ref{fig:optical_orb}),
the magnitude of the peak at $\omega_1$ increased significantly.
So the onset of this low-energy peak
at $\omega_1$ is mostly due to antiferromagnetism, which opens up a gap
at the chemical potential and shifts part of the Drude weight to the region around $\omega_1$. On the other hand, the high-energy peak at $\omega_2$ comes from
the interband transitions that are already present in the original
multi-orbital model. However, we cannot distinguish along which
direction the optical conductivity dominates since there are multiple
crossings (as a function of frequency) between the two curves of $\sigma_{xx}$ and
$\sigma_{yy}$. But, naively, $\sigma_{xx}$ does have higher values in
most of the regions around $\omega_1$, which matches the experimental
results \cite{Dusza2011} at the low-temperature antiferromagnetic
phase. Also of note is the region around $\omega_1$
where $\sigma_{xx}$ displays a single peak, whereas $\sigma_{yy}$
acquires a double-peak structure, which is also reproduced in a recent study \cite{Yin2011} using the combination of density functional theory and dynamical mean-field theory (DFT+DMFT). Anisotropy of the Drude weight is presented as a
function of $U$ and $\Delta$ in Fig.~\ref{fig:optical_mag_orb}(b). If
we assume an isotropic scattering rate, the dc conductivity anisotropy
has the correct directionality in a very large region of parameter
space. However, considering the $(\pi,0,\pi)$ antiferromagnetism, we
should expect a larger scattering rate along the $x$ direction, where
the spins are antiparallel to their neighbors. This should further
restrict the parameter space of interest.  Nonetheless, we find that
in a reasonable range of parameters, both orbital and magnetic order
underlie the resistivity anisotropy in a multi-orbital
model of the pnictides.

\section{Final remarks}
\label{sec:con}

In summary, we have solved a three-dimensional five-orbital model
using the mean-field approximation in the presence of both orbital and
magnetic order.  We modeled the orbital order phenomenologically as an
energy splitting term between the two otherwise degenerate $d_{xz}$
and $d_{yz}$ orbitals, and find that it is solely responsible for the
splitting between the bands at $X$ and $Y$ observed by polarized
ARPES.\cite{Yi2011} This orbital nematic order also causes the Drude
weight difference between the two diagonal directions for the
paramagnetic phase, in agreement with the analysis based on optical
conductivity.\cite{Dusza2011}  On the other hand, the magnetic order
sets in at a lower temperature. It opens up a gap at the Fermi energy and
shifts part of the Drude weight to high energies, leading to the
emergence of a peak structure at $\omega_1 \approx 0.2$ in the optical
conductivity. However, magnetic order alone cannot explain the
anisotropic dispersions, especially the band splitting between $X$ and
$Y$. Thus both orbital and magnetic orders are present in the
low-temperature antiferromagnetic phase, and together induce various
anisotropies seen by different experimental techniques. This result is
in contrast with earlier studies,
\cite{Daghofer2010a,Bascones2010,Valenzuela2010} which claimed that
the magnetic order is the driving mechanism for the anisotropy. Thus,
our study establishes the importance of orbital order, irrespective of
whether long-range magnetic order is present or not.

Besides orbital order, the anisotropy observed above the
magnetostructural transition can come from a different mechanism, for
example, static short-range magnetic order. Its existence is indeed supported by recent INS experiments.\cite{Harriger2011} In principle, we can model this by a probability distribution of the ordering wave vector peaked at $\bm{q}=(\pi,0,\pi)$,\cite{Harrison2007} and investigate its modification to the electronic structure. It is expected that the magnitude of the subpeak at $\omega_1$ in Fig.~\ref{fig:optical_orb}(b) will increase due to this short-range magnetic order, thus yielding better agreement with experiments.\cite{Dusza2011} Nevertheless, the orbital order should always be present until the band splitting vanishes.

Of course, it remains an outstanding issue of whether the nematic order
above the magnetostructural transition is orbitally or
magnetically driven. Thermodynamic measurement cannot distinguish them as both transitions fall into the Ising universality class. Indeed, increasing sample quality does drive the magnetic transition closer to the structural transition.\cite{Jesche2010}
In our point of view, these two degrees of freedom actually
coexist and cooperate with one another. A nonzero orbital order can
certainly induce a spin nematicity without long-range magnetic order,
and vice versa. It will be interesting if future INS experiments can
settle the onset temperature of the spin nematic order and compare it
to that of the orbital order measured by ARPES.\cite{Yi2011} However,
to account for the large anisotropy of the magnetic exchanges,\cite{Zhao2009,Harriger2011} orbital degrees of freedom have to be considered.

One ingredient we ignored in our theory is the role of the
structural distortions, which in principle should enhance the orbital order.\cite{Liu2011}
We also neglected possible strong-correlation physics, which may cause an orbital-selective Mott transition.\cite{Laad2011}
To conclude, orbital order is an important part of the minimal model of iron-based superconductors. It induces strong exchange anisotropy between the local moments, and leads to an electron nematic order on the itinerant level, governing the physics over a very large energy scale. The study of the interplay between the orbital and magnetic orders should shed new insight into the mechanism of superconductivity in this multi-orbital system.

\begin{acknowledgments}
We would like to acknowledge helpful discussions with C.-C.~Chen, J.-H.~Chu, L.~Degiorgi, D.~S.~Dessau, S.~Hong, W.-C.~Lee, D.~H.~Lu, Q.~Wang, and M.~Yi, especially C.-C.~Chen for his suggestion of Ref.~\onlinecite{Graser2010}.
This work is supported by the NSF under Grant No.~DMR-0940992 and the Center for
Emergent Superconductivity, a DOE Energy Frontier Research Center, Grant No.~DE-AC0298CH1088.
\end{acknowledgments}

\bibliography{feas_mf}

\begin{thebibliography}{61}%
\makeatletter
\providecommand \@ifxundefined [1]{%
 \@ifx{#1\undefined}
}%
\providecommand \@ifnum [1]{%
 \ifnum #1\expandafter \@firstoftwo
 \else \expandafter \@secondoftwo
 \fi
}%
\providecommand \@ifx [1]{%
 \ifx #1\expandafter \@firstoftwo
 \else \expandafter \@secondoftwo
 \fi
}%
\providecommand \natexlab [1]{#1}%
\providecommand \enquote  [1]{``#1''}%
\providecommand \bibnamefont  [1]{#1}%
\providecommand \bibfnamefont [1]{#1}%
\providecommand \citenamefont [1]{#1}%
\providecommand \href@noop [0]{\@secondoftwo}%
\providecommand \href [0]{\begingroup \@sanitize@url \@href}%
\providecommand \@href[1]{\@@startlink{#1}\@@href}%
\providecommand \@@href[1]{\endgroup#1\@@endlink}%
\providecommand \@sanitize@url [0]{\catcode `\\12\catcode `\$12\catcode
  `\&12\catcode `\#12\catcode `\^12\catcode `\_12\catcode `\%12\relax}%
\providecommand \@@startlink[1]{}%
\providecommand \@@endlink[0]{}%
\providecommand \url  [0]{\begingroup\@sanitize@url \@url }%
\providecommand \@url [1]{\endgroup\@href {#1}{\urlprefix }}%
\providecommand \urlprefix  [0]{URL }%
\providecommand \Eprint [0]{\href }%
\providecommand \doibase [0]{http://dx.doi.org/}%
\providecommand \selectlanguage [0]{\@gobble}%
\providecommand \bibinfo  [0]{\@secondoftwo}%
\providecommand \bibfield  [0]{\@secondoftwo}%
\providecommand \translation [1]{[#1]}%
\providecommand \BibitemOpen [0]{}%
\providecommand \bibitemStop [0]{}%
\providecommand \bibitemNoStop [0]{.\EOS\space}%
\providecommand \EOS [0]{\spacefactor3000\relax}%
\providecommand \BibitemShut  [1]{\csname bibitem#1\endcsname}%
\let\auto@bib@innerbib\@empty
\bibitem [{\citenamefont {Zhao}\ \emph {et~al.}(2009)\citenamefont {Zhao},
  \citenamefont {Adroja}, \citenamefont {Yao}, \citenamefont {Bewley},
  \citenamefont {Li}, \citenamefont {Wang}, \citenamefont {Wu}, \citenamefont
  {Chen}, \citenamefont {Hu},\ and\ \citenamefont {Dai}}]{Zhao2009}%
  \BibitemOpen
  \bibfield  {author} {\bibinfo {author} {\bibfnamefont {J.}~\bibnamefont
  {Zhao}}, \bibinfo {author} {\bibfnamefont {D.~T.}\ \bibnamefont {Adroja}},
  \bibinfo {author} {\bibfnamefont {D.-X.}\ \bibnamefont {Yao}}, \bibinfo
  {author} {\bibfnamefont {R.}~\bibnamefont {Bewley}}, \bibinfo {author}
  {\bibfnamefont {S.}~\bibnamefont {Li}}, \bibinfo {author} {\bibfnamefont
  {X.~F.}\ \bibnamefont {Wang}}, \bibinfo {author} {\bibfnamefont
  {G.}~\bibnamefont {Wu}}, \bibinfo {author} {\bibfnamefont {X.~H.}\
  \bibnamefont {Chen}}, \bibinfo {author} {\bibfnamefont {J.}~\bibnamefont
  {Hu}}, \ and\ \bibinfo {author} {\bibfnamefont {P.}~\bibnamefont {Dai}},\
  }\href@noop {} {\bibfield  {journal} {\bibinfo  {journal} {Nature Physics}\
  }\textbf {\bibinfo {volume} {5}},\ \bibinfo {pages} {555} (\bibinfo {year}
  {2009})}\BibitemShut {NoStop}%
\bibitem [{\citenamefont {Shimojima}\ \emph {et~al.}(2010)\citenamefont
  {Shimojima}, \citenamefont {Ishizaka}, \citenamefont {Ishida}, \citenamefont
  {Katayama}, \citenamefont {Ohgushi}, \citenamefont {Kiss}, \citenamefont
  {Okawa}, \citenamefont {Togashi}, \citenamefont {Wang}, \citenamefont {Chen},
  \citenamefont {Watanabe}, \citenamefont {Kadota}, \citenamefont {Oguchi},
  \citenamefont {Chainani},\ and\ \citenamefont {Shin}}]{Shimojima2010}%
  \BibitemOpen
  \bibfield  {author} {\bibinfo {author} {\bibfnamefont {T.}~\bibnamefont
  {Shimojima}}, \bibinfo {author} {\bibfnamefont {K.}~\bibnamefont {Ishizaka}},
  \bibinfo {author} {\bibfnamefont {Y.}~\bibnamefont {Ishida}}, \bibinfo
  {author} {\bibfnamefont {N.}~\bibnamefont {Katayama}}, \bibinfo {author}
  {\bibfnamefont {K.}~\bibnamefont {Ohgushi}}, \bibinfo {author} {\bibfnamefont
  {T.}~\bibnamefont {Kiss}}, \bibinfo {author} {\bibfnamefont {M.}~\bibnamefont
  {Okawa}}, \bibinfo {author} {\bibfnamefont {T.}~\bibnamefont {Togashi}},
  \bibinfo {author} {\bibfnamefont {X.-Y.}\ \bibnamefont {Wang}}, \bibinfo
  {author} {\bibfnamefont {C.-T.}\ \bibnamefont {Chen}}, \bibinfo {author}
  {\bibfnamefont {S.}~\bibnamefont {Watanabe}}, \bibinfo {author}
  {\bibfnamefont {R.}~\bibnamefont {Kadota}}, \bibinfo {author} {\bibfnamefont
  {T.}~\bibnamefont {Oguchi}}, \bibinfo {author} {\bibfnamefont
  {A.}~\bibnamefont {Chainani}}, \ and\ \bibinfo {author} {\bibfnamefont
  {S.}~\bibnamefont {Shin}},\ }\href@noop {} {\bibfield  {journal} {\bibinfo
  {journal} {Phys. Rev. Lett.}\ }\textbf {\bibinfo {volume} {104}},\ \bibinfo
  {pages} {057002} (\bibinfo {year} {2010})}\BibitemShut {NoStop}%
\bibitem [{\citenamefont {Wang}\ \emph {et~al.}()\citenamefont {Wang},
  \citenamefont {Sun}, \citenamefont {Rotenberg}, \citenamefont {Ronning},
  \citenamefont {Bauer}, \citenamefont {Lin}, \citenamefont {Markiewicz},
  \citenamefont {Lindroos}, \citenamefont {Barbiellini}, \citenamefont
  {Bansil},\ and\ \citenamefont {Dessau}}]{Wang2010}%
  \BibitemOpen
  \bibfield  {author} {\bibinfo {author} {\bibfnamefont {Q.}~\bibnamefont
  {Wang}}, \bibinfo {author} {\bibfnamefont {Z.}~\bibnamefont {Sun}}, \bibinfo
  {author} {\bibfnamefont {E.}~\bibnamefont {Rotenberg}}, \bibinfo {author}
  {\bibfnamefont {F.}~\bibnamefont {Ronning}}, \bibinfo {author} {\bibfnamefont
  {E.~D.}\ \bibnamefont {Bauer}}, \bibinfo {author} {\bibfnamefont
  {H.}~\bibnamefont {Lin}}, \bibinfo {author} {\bibfnamefont {R.~S.}\
  \bibnamefont {Markiewicz}}, \bibinfo {author} {\bibfnamefont
  {M.}~\bibnamefont {Lindroos}}, \bibinfo {author} {\bibfnamefont
  {B.}~\bibnamefont {Barbiellini}}, \bibinfo {author} {\bibfnamefont
  {A.}~\bibnamefont {Bansil}}, \ and\ \bibinfo {author} {\bibfnamefont {D.~S.}\
  \bibnamefont {Dessau}},\ }\href@noop {} {\ }\Eprint
  {http://arxiv.org/abs/arXiv:1009.0271} {arXiv:1009.0271} \BibitemShut
  {NoStop}%
\bibitem [{\citenamefont {Chuang}\ \emph {et~al.}(2010)\citenamefont {Chuang},
  \citenamefont {Allan}, \citenamefont {Lee}, \citenamefont {Xie},
  \citenamefont {Ni}, \citenamefont {Bud'ko}, \citenamefont {Boebinger},
  \citenamefont {Canfield},\ and\ \citenamefont {Davis}}]{Chuang2010}%
  \BibitemOpen
  \bibfield  {author} {\bibinfo {author} {\bibfnamefont {T.-M.}\ \bibnamefont
  {Chuang}}, \bibinfo {author} {\bibfnamefont {M.~P.}\ \bibnamefont {Allan}},
  \bibinfo {author} {\bibfnamefont {J.}~\bibnamefont {Lee}}, \bibinfo {author}
  {\bibfnamefont {Y.}~\bibnamefont {Xie}}, \bibinfo {author} {\bibfnamefont
  {N.}~\bibnamefont {Ni}}, \bibinfo {author} {\bibfnamefont {S.~L.}\
  \bibnamefont {Bud'ko}}, \bibinfo {author} {\bibfnamefont {G.~S.}\
  \bibnamefont {Boebinger}}, \bibinfo {author} {\bibfnamefont {P.~C.}\
  \bibnamefont {Canfield}}, \ and\ \bibinfo {author} {\bibfnamefont {J.~C.}\
  \bibnamefont {Davis}},\ }\href@noop {} {\bibfield  {journal} {\bibinfo
  {journal} {Science}\ }\textbf {\bibinfo {volume} {327}},\ \bibinfo {pages}
  {181} (\bibinfo {year} {2010})}\BibitemShut {NoStop}%
\bibitem [{\citenamefont {Chu}\ \emph {et~al.}(2010)\citenamefont {Chu},
  \citenamefont {Analytis}, \citenamefont {De~Greve}, \citenamefont {McMahon},
  \citenamefont {Islam}, \citenamefont {Yamamoto},\ and\ \citenamefont
  {Fisher}}]{Chu2010}%
  \BibitemOpen
  \bibfield  {author} {\bibinfo {author} {\bibfnamefont {J.-H.}\ \bibnamefont
  {Chu}}, \bibinfo {author} {\bibfnamefont {J.~G.}\ \bibnamefont {Analytis}},
  \bibinfo {author} {\bibfnamefont {K.}~\bibnamefont {De~Greve}}, \bibinfo
  {author} {\bibfnamefont {P.~L.}\ \bibnamefont {McMahon}}, \bibinfo {author}
  {\bibfnamefont {Z.}~\bibnamefont {Islam}}, \bibinfo {author} {\bibfnamefont
  {Y.}~\bibnamefont {Yamamoto}}, \ and\ \bibinfo {author} {\bibfnamefont
  {I.~R.}\ \bibnamefont {Fisher}},\ }\href@noop {} {\bibfield  {journal}
  {\bibinfo  {journal} {Science}\ }\textbf {\bibinfo {volume} {329}},\ \bibinfo
  {pages} {824} (\bibinfo {year} {2010})}\BibitemShut {NoStop}%
\bibitem [{\citenamefont {Tanatar}\ \emph {et~al.}(2010)\citenamefont
  {Tanatar}, \citenamefont {Blomberg}, \citenamefont {Kreyssig}, \citenamefont
  {Kim}, \citenamefont {Ni}, \citenamefont {Thaler}, \citenamefont {Bud'ko},
  \citenamefont {Canfield}, \citenamefont {Goldman}, \citenamefont {Mazin},\
  and\ \citenamefont {Prozorov}}]{Tanatar2010}%
  \BibitemOpen
  \bibfield  {author} {\bibinfo {author} {\bibfnamefont {M.~A.}\ \bibnamefont
  {Tanatar}}, \bibinfo {author} {\bibfnamefont {E.~C.}\ \bibnamefont
  {Blomberg}}, \bibinfo {author} {\bibfnamefont {A.}~\bibnamefont {Kreyssig}},
  \bibinfo {author} {\bibfnamefont {M.~G.}\ \bibnamefont {Kim}}, \bibinfo
  {author} {\bibfnamefont {N.}~\bibnamefont {Ni}}, \bibinfo {author}
  {\bibfnamefont {A.}~\bibnamefont {Thaler}}, \bibinfo {author} {\bibfnamefont
  {S.~L.}\ \bibnamefont {Bud'ko}}, \bibinfo {author} {\bibfnamefont {P.~C.}\
  \bibnamefont {Canfield}}, \bibinfo {author} {\bibfnamefont {A.~I.}\
  \bibnamefont {Goldman}}, \bibinfo {author} {\bibfnamefont {I.~I.}\
  \bibnamefont {Mazin}}, \ and\ \bibinfo {author} {\bibfnamefont
  {R.}~\bibnamefont {Prozorov}},\ }\href@noop {} {\bibfield  {journal}
  {\bibinfo  {journal} {Phys. Rev. B}\ }\textbf {\bibinfo {volume} {81}},\
  \bibinfo {pages} {184508} (\bibinfo {year} {2010})}\BibitemShut {NoStop}%
\bibitem [{\citenamefont {Kuo}\ \emph {et~al.}(2011)\citenamefont {Kuo},
  \citenamefont {Chu}, \citenamefont {Riggs}, \citenamefont {Yu}, \citenamefont
  {McMahon}, \citenamefont {De~Greve}, \citenamefont {Yamamoto}, \citenamefont
  {Analytis},\ and\ \citenamefont {Fisher}}]{Kuo2011}%
  \BibitemOpen
  \bibfield  {author} {\bibinfo {author} {\bibfnamefont {H.-H.}\ \bibnamefont
  {Kuo}}, \bibinfo {author} {\bibfnamefont {J.-H.}\ \bibnamefont {Chu}},
  \bibinfo {author} {\bibfnamefont {S.~C.}\ \bibnamefont {Riggs}}, \bibinfo
  {author} {\bibfnamefont {L.}~\bibnamefont {Yu}}, \bibinfo {author}
  {\bibfnamefont {P.~L.}\ \bibnamefont {McMahon}}, \bibinfo {author}
  {\bibfnamefont {K.}~\bibnamefont {De~Greve}}, \bibinfo {author}
  {\bibfnamefont {Y.}~\bibnamefont {Yamamoto}}, \bibinfo {author}
  {\bibfnamefont {J.~G.}\ \bibnamefont {Analytis}}, \ and\ \bibinfo {author}
  {\bibfnamefont {I.~R.}\ \bibnamefont {Fisher}},\ }\href@noop {} {\bibfield
  {journal} {\bibinfo  {journal} {Phys. Rev. B}\ }\textbf {\bibinfo {volume}
  {84}},\ \bibinfo {pages} {054540} (\bibinfo {year} {2011})}\BibitemShut
  {NoStop}%
\bibitem [{\citenamefont {Ying}\ \emph {et~al.}(2011)\citenamefont {Ying},
  \citenamefont {Wang}, \citenamefont {Wu}, \citenamefont {Xiang},
  \citenamefont {Liu}, \citenamefont {Yan}, \citenamefont {Wang}, \citenamefont
  {Zhang}, \citenamefont {Ye}, \citenamefont {Cheng}, \citenamefont {Hu},\ and\
  \citenamefont {Chen}}]{Ying2011}%
  \BibitemOpen
  \bibfield  {author} {\bibinfo {author} {\bibfnamefont {J.~J.}\ \bibnamefont
  {Ying}}, \bibinfo {author} {\bibfnamefont {X.~F.}\ \bibnamefont {Wang}},
  \bibinfo {author} {\bibfnamefont {T.}~\bibnamefont {Wu}}, \bibinfo {author}
  {\bibfnamefont {Z.~J.}\ \bibnamefont {Xiang}}, \bibinfo {author}
  {\bibfnamefont {R.~H.}\ \bibnamefont {Liu}}, \bibinfo {author} {\bibfnamefont
  {Y.~J.}\ \bibnamefont {Yan}}, \bibinfo {author} {\bibfnamefont {A.~F.}\
  \bibnamefont {Wang}}, \bibinfo {author} {\bibfnamefont {M.}~\bibnamefont
  {Zhang}}, \bibinfo {author} {\bibfnamefont {G.~J.}\ \bibnamefont {Ye}},
  \bibinfo {author} {\bibfnamefont {P.}~\bibnamefont {Cheng}}, \bibinfo
  {author} {\bibfnamefont {J.~P.}\ \bibnamefont {Hu}}, \ and\ \bibinfo {author}
  {\bibfnamefont {X.~H.}\ \bibnamefont {Chen}},\ }\href@noop {} {\bibfield
  {journal} {\bibinfo  {journal} {Phys. Rev. Lett.}\ }\textbf {\bibinfo
  {volume} {107}},\ \bibinfo {pages} {067001} (\bibinfo {year}
  {2011})}\BibitemShut {NoStop}%
\bibitem [{\citenamefont {Dusza}\ \emph {et~al.}(2011)\citenamefont {Dusza},
  \citenamefont {Lucarelli}, \citenamefont {Pfuner}, \citenamefont {Chu},
  \citenamefont {Fisher},\ and\ \citenamefont {Degiorgi}}]{Dusza2011}%
  \BibitemOpen
  \bibfield  {author} {\bibinfo {author} {\bibfnamefont {A.}~\bibnamefont
  {Dusza}}, \bibinfo {author} {\bibfnamefont {A.}~\bibnamefont {Lucarelli}},
  \bibinfo {author} {\bibfnamefont {F.}~\bibnamefont {Pfuner}}, \bibinfo
  {author} {\bibfnamefont {J.-H.}\ \bibnamefont {Chu}}, \bibinfo {author}
  {\bibfnamefont {I.~R.}\ \bibnamefont {Fisher}}, \ and\ \bibinfo {author}
  {\bibfnamefont {L.}~\bibnamefont {Degiorgi}},\ }\href@noop {} {\bibfield
  {journal} {\bibinfo  {journal} {Europhys. Lett.}\ }\textbf {\bibinfo {volume}
  {93}},\ \bibinfo {pages} {37002} (\bibinfo {year} {2011})}\BibitemShut
  {NoStop}%
\bibitem [{\citenamefont {Nakajima}\ \emph {et~al.}(2011)\citenamefont
  {Nakajima}, \citenamefont {Liang}, \citenamefont {Ishida}, \citenamefont
  {Tomioka}, \citenamefont {Kihou}, \citenamefont {Lee}, \citenamefont {Iyo},
  \citenamefont {Eisaki}, \citenamefont {Kakeshita}, \citenamefont {Ito},\ and\
  \citenamefont {Uchida}}]{Nakajima2011}%
  \BibitemOpen
  \bibfield  {author} {\bibinfo {author} {\bibfnamefont {M.}~\bibnamefont
  {Nakajima}}, \bibinfo {author} {\bibfnamefont {T.}~\bibnamefont {Liang}},
  \bibinfo {author} {\bibfnamefont {S.}~\bibnamefont {Ishida}}, \bibinfo
  {author} {\bibfnamefont {Y.}~\bibnamefont {Tomioka}}, \bibinfo {author}
  {\bibfnamefont {K.}~\bibnamefont {Kihou}}, \bibinfo {author} {\bibfnamefont
  {C.~H.}\ \bibnamefont {Lee}}, \bibinfo {author} {\bibfnamefont
  {A.}~\bibnamefont {Iyo}}, \bibinfo {author} {\bibfnamefont {H.}~\bibnamefont
  {Eisaki}}, \bibinfo {author} {\bibfnamefont {T.}~\bibnamefont {Kakeshita}},
  \bibinfo {author} {\bibfnamefont {T.}~\bibnamefont {Ito}}, \ and\ \bibinfo
  {author} {\bibfnamefont {S.}~\bibnamefont {Uchida}},\ }\href@noop {}
  {\bibfield  {journal} {\bibinfo  {journal} {Proceedings of the National
  Academy of Sciences}\ }\textbf {\bibinfo {volume} {108}},\ \bibinfo {pages}
  {12238} (\bibinfo {year} {2011})}\BibitemShut {NoStop}%
\bibitem [{\citenamefont {Lucarelli}\ \emph {et~al.}()\citenamefont
  {Lucarelli}, \citenamefont {Dusza}, \citenamefont {Sanna}, \citenamefont
  {Massidda}, \citenamefont {Chu}, \citenamefont {Fisher},\ and\ \citenamefont
  {Degiorgi}}]{Lucarelli2011}%
  \BibitemOpen
  \bibfield  {author} {\bibinfo {author} {\bibfnamefont {A.}~\bibnamefont
  {Lucarelli}}, \bibinfo {author} {\bibfnamefont {A.}~\bibnamefont {Dusza}},
  \bibinfo {author} {\bibfnamefont {A.}~\bibnamefont {Sanna}}, \bibinfo
  {author} {\bibfnamefont {S.}~\bibnamefont {Massidda}}, \bibinfo {author}
  {\bibfnamefont {J.~H.}\ \bibnamefont {Chu}}, \bibinfo {author} {\bibfnamefont
  {I.~R.}\ \bibnamefont {Fisher}}, \ and\ \bibinfo {author} {\bibfnamefont
  {L.}~\bibnamefont {Degiorgi}},\ }\href@noop {} {\ }\Eprint
  {http://arxiv.org/abs/arXiv:1107.0670} {arXiv:1107.0670} \BibitemShut
  {NoStop}%
\bibitem [{\citenamefont {Yi}\ \emph {et~al.}(2011)\citenamefont {Yi},
  \citenamefont {Lu}, \citenamefont {Chu}, \citenamefont {Analytis},
  \citenamefont {Sorini}, \citenamefont {Kemper}, \citenamefont {Moritz},
  \citenamefont {Mo}, \citenamefont {Moore}, \citenamefont {Hashimoto},
  \citenamefont {Lee}, \citenamefont {Hussain}, \citenamefont {Devereaux},
  \citenamefont {Fisher},\ and\ \citenamefont {Shen}}]{Yi2011}%
  \BibitemOpen
  \bibfield  {author} {\bibinfo {author} {\bibfnamefont {M.}~\bibnamefont
  {Yi}}, \bibinfo {author} {\bibfnamefont {D.}~\bibnamefont {Lu}}, \bibinfo
  {author} {\bibfnamefont {J.-H.}\ \bibnamefont {Chu}}, \bibinfo {author}
  {\bibfnamefont {J.~G.}\ \bibnamefont {Analytis}}, \bibinfo {author}
  {\bibfnamefont {A.~P.}\ \bibnamefont {Sorini}}, \bibinfo {author}
  {\bibfnamefont {A.~F.}\ \bibnamefont {Kemper}}, \bibinfo {author}
  {\bibfnamefont {B.}~\bibnamefont {Moritz}}, \bibinfo {author} {\bibfnamefont
  {S.-K.}\ \bibnamefont {Mo}}, \bibinfo {author} {\bibfnamefont {R.~G.}\
  \bibnamefont {Moore}}, \bibinfo {author} {\bibfnamefont {M.}~\bibnamefont
  {Hashimoto}}, \bibinfo {author} {\bibfnamefont {W.-S.}\ \bibnamefont {Lee}},
  \bibinfo {author} {\bibfnamefont {Z.}~\bibnamefont {Hussain}}, \bibinfo
  {author} {\bibfnamefont {T.~P.}\ \bibnamefont {Devereaux}}, \bibinfo {author}
  {\bibfnamefont {I.~R.}\ \bibnamefont {Fisher}}, \ and\ \bibinfo {author}
  {\bibfnamefont {Z.-X.}\ \bibnamefont {Shen}},\ }\href@noop {} {\bibfield
  {journal} {\bibinfo  {journal} {Proceedings of the National Academy of
  Sciences}\ }\textbf {\bibinfo {volume} {108}},\ \bibinfo {pages} {6878}
  (\bibinfo {year} {2011})}\BibitemShut {NoStop}%
\bibitem [{\citenamefont {Fisher}\ \emph {et~al.}(2011)\citenamefont {Fisher},
  \citenamefont {Degiorgi},\ and\ \citenamefont {Shen}}]{Fisher2011}%
  \BibitemOpen
  \bibfield  {author} {\bibinfo {author} {\bibfnamefont {I.~R.}\ \bibnamefont
  {Fisher}}, \bibinfo {author} {\bibfnamefont {L.}~\bibnamefont {Degiorgi}}, \
  and\ \bibinfo {author} {\bibfnamefont {Z.~X.}\ \bibnamefont {Shen}},\
  }\href@noop {} {\bibfield  {journal} {\bibinfo  {journal} {Reports on
  Progress in Physics}\ }\textbf {\bibinfo {volume} {74}},\ \bibinfo {pages}
  {124506} (\bibinfo {year} {2011})}\BibitemShut {NoStop}%
\bibitem [{\citenamefont {Wu}\ \emph {et~al.}(2008)\citenamefont {Wu},
  \citenamefont {Phillips},\ and\ \citenamefont {Neto}}]{Wu2008}%
  \BibitemOpen
  \bibfield  {author} {\bibinfo {author} {\bibfnamefont {J.}~\bibnamefont
  {Wu}}, \bibinfo {author} {\bibfnamefont {P.}~\bibnamefont {Phillips}}, \ and\
  \bibinfo {author} {\bibfnamefont {A.~H.~C.}\ \bibnamefont {Neto}},\
  }\href@noop {} {\bibfield  {journal} {\bibinfo  {journal} {Phys. Rev. Lett.}\
  }\textbf {\bibinfo {volume} {101}},\ \bibinfo {eid} {126401} (\bibinfo {year}
  {2008})}\BibitemShut {NoStop}%
\bibitem [{\citenamefont {Dai}\ \emph {et~al.}(2009)\citenamefont {Dai},
  \citenamefont {Si}, \citenamefont {Zhu},\ and\ \citenamefont
  {Abrahams}}]{Dai2009}%
  \BibitemOpen
  \bibfield  {author} {\bibinfo {author} {\bibfnamefont {J.}~\bibnamefont
  {Dai}}, \bibinfo {author} {\bibfnamefont {Q.}~\bibnamefont {Si}}, \bibinfo
  {author} {\bibfnamefont {J.-X.}\ \bibnamefont {Zhu}}, \ and\ \bibinfo
  {author} {\bibfnamefont {E.}~\bibnamefont {Abrahams}},\ }\href@noop {}
  {\bibfield  {journal} {\bibinfo  {journal} {Proceedings of the National
  Academy of Sciences}\ }\textbf {\bibinfo {volume} {106}},\ \bibinfo {pages}
  {4118} (\bibinfo {year} {2009})}\BibitemShut {NoStop}%
\bibitem [{\citenamefont {Kou}\ \emph {et~al.}(2009)\citenamefont {Kou},
  \citenamefont {Li},\ and\ \citenamefont {Weng}}]{Kou2009}%
  \BibitemOpen
  \bibfield  {author} {\bibinfo {author} {\bibfnamefont {S.-P.}\ \bibnamefont
  {Kou}}, \bibinfo {author} {\bibfnamefont {T.}~\bibnamefont {Li}}, \ and\
  \bibinfo {author} {\bibfnamefont {Z.-Y.}\ \bibnamefont {Weng}},\ }\href@noop
  {} {\bibfield  {journal} {\bibinfo  {journal} {Europhys. Lett.}\ }\textbf
  {\bibinfo {volume} {88}},\ \bibinfo {pages} {17010} (\bibinfo {year}
  {2009})}\BibitemShut {NoStop}%
\bibitem [{\citenamefont {Kr\"{u}ger}\ \emph {et~al.}(2009)\citenamefont
  {Kr\"{u}ger}, \citenamefont {Kumar}, \citenamefont {Zaanen},\ and\
  \citenamefont {van~den Brink}}]{Kruger2009}%
  \BibitemOpen
  \bibfield  {author} {\bibinfo {author} {\bibfnamefont {F.}~\bibnamefont
  {Kr\"{u}ger}}, \bibinfo {author} {\bibfnamefont {S.}~\bibnamefont {Kumar}},
  \bibinfo {author} {\bibfnamefont {J.}~\bibnamefont {Zaanen}}, \ and\ \bibinfo
  {author} {\bibfnamefont {J.}~\bibnamefont {van~den Brink}},\ }\href@noop {}
  {\bibfield  {journal} {\bibinfo  {journal} {Phys. Rev. B}\ }\textbf {\bibinfo
  {volume} {79}},\ \bibinfo {eid} {054504} (\bibinfo {year}
  {2009})}\BibitemShut {NoStop}%
\bibitem [{\citenamefont {Lv}\ \emph {et~al.}(2009)\citenamefont {Lv},
  \citenamefont {Wu},\ and\ \citenamefont {Phillips}}]{Lv2009}%
  \BibitemOpen
  \bibfield  {author} {\bibinfo {author} {\bibfnamefont {W.}~\bibnamefont
  {Lv}}, \bibinfo {author} {\bibfnamefont {J.}~\bibnamefont {Wu}}, \ and\
  \bibinfo {author} {\bibfnamefont {P.}~\bibnamefont {Phillips}},\ }\href@noop
  {} {\bibfield  {journal} {\bibinfo  {journal} {Phys. Rev. B}\ }\textbf
  {\bibinfo {volume} {80}},\ \bibinfo {pages} {224506} (\bibinfo {year}
  {2009})}\BibitemShut {NoStop}%
\bibitem [{\citenamefont {Lee}\ \emph {et~al.}(2009)\citenamefont {Lee},
  \citenamefont {Yin},\ and\ \citenamefont {Ku}}]{Lee2009}%
  \BibitemOpen
  \bibfield  {author} {\bibinfo {author} {\bibfnamefont {C.-C.}\ \bibnamefont
  {Lee}}, \bibinfo {author} {\bibfnamefont {W.-G.}\ \bibnamefont {Yin}}, \ and\
  \bibinfo {author} {\bibfnamefont {W.}~\bibnamefont {Ku}},\ }\href@noop {}
  {\bibfield  {journal} {\bibinfo  {journal} {Phys. Rev. Lett.}\ }\textbf
  {\bibinfo {volume} {103}},\ \bibinfo {pages} {267001} (\bibinfo {year}
  {2009})}\BibitemShut {NoStop}%
\bibitem [{\citenamefont {Chen}\ \emph {et~al.}(2009)\citenamefont {Chen},
  \citenamefont {Moritz}, \citenamefont {van~den Brink}, \citenamefont
  {Devereaux},\ and\ \citenamefont {Singh}}]{Chen2009}%
  \BibitemOpen
  \bibfield  {author} {\bibinfo {author} {\bibfnamefont {C.-C.}\ \bibnamefont
  {Chen}}, \bibinfo {author} {\bibfnamefont {B.}~\bibnamefont {Moritz}},
  \bibinfo {author} {\bibfnamefont {J.}~\bibnamefont {van~den Brink}}, \bibinfo
  {author} {\bibfnamefont {T.~P.}\ \bibnamefont {Devereaux}}, \ and\ \bibinfo
  {author} {\bibfnamefont {R.~R.~P.}\ \bibnamefont {Singh}},\ }\href@noop {}
  {\bibfield  {journal} {\bibinfo  {journal} {Phys. Rev. B}\ }\textbf {\bibinfo
  {volume} {80}},\ \bibinfo {pages} {180418} (\bibinfo {year}
  {2009})}\BibitemShut {NoStop}%
\bibitem [{\citenamefont {Kugel}\ and\ \citenamefont
  {Khomskii}(1982)}]{Kugel1982}%
  \BibitemOpen
  \bibfield  {author} {\bibinfo {author} {\bibfnamefont {K.~I.}\ \bibnamefont
  {Kugel}}\ and\ \bibinfo {author} {\bibfnamefont {D.~I.}\ \bibnamefont
  {Khomskii}},\ }\href@noop {} {\bibfield  {journal} {\bibinfo  {journal}
  {Soviet Physics Uspekhi}\ }\textbf {\bibinfo {volume} {25}},\ \bibinfo
  {pages} {231} (\bibinfo {year} {1982})}\BibitemShut {NoStop}%
\bibitem [{\citenamefont {Lv}\ \emph {et~al.}(2010)\citenamefont {Lv},
  \citenamefont {Kr\"uger},\ and\ \citenamefont {Phillips}}]{Lv2010}%
  \BibitemOpen
  \bibfield  {author} {\bibinfo {author} {\bibfnamefont {W.}~\bibnamefont
  {Lv}}, \bibinfo {author} {\bibfnamefont {F.}~\bibnamefont {Kr\"uger}}, \ and\
  \bibinfo {author} {\bibfnamefont {P.}~\bibnamefont {Phillips}},\ }\href@noop
  {} {\bibfield  {journal} {\bibinfo  {journal} {Phys. Rev. B}\ }\textbf
  {\bibinfo {volume} {82}},\ \bibinfo {pages} {045125} (\bibinfo {year}
  {2010})}\BibitemShut {NoStop}%
\bibitem [{\citenamefont {Yin}\ \emph {et~al.}(2010)\citenamefont {Yin},
  \citenamefont {Lee},\ and\ \citenamefont {Ku}}]{Yin2010}%
  \BibitemOpen
  \bibfield  {author} {\bibinfo {author} {\bibfnamefont {W.-G.}\ \bibnamefont
  {Yin}}, \bibinfo {author} {\bibfnamefont {C.-C.}\ \bibnamefont {Lee}}, \ and\
  \bibinfo {author} {\bibfnamefont {W.}~\bibnamefont {Ku}},\ }\href@noop {}
  {\bibfield  {journal} {\bibinfo  {journal} {Phys. Rev. Lett.}\ }\textbf
  {\bibinfo {volume} {105}},\ \bibinfo {pages} {107004} (\bibinfo {year}
  {2010})}\BibitemShut {NoStop}%
\bibitem [{\citenamefont {Harriger}\ \emph {et~al.}(2011)\citenamefont
  {Harriger}, \citenamefont {Luo}, \citenamefont {Liu}, \citenamefont {Frost},
  \citenamefont {Hu}, \citenamefont {Norman},\ and\ \citenamefont
  {Dai}}]{Harriger2011}%
  \BibitemOpen
  \bibfield  {author} {\bibinfo {author} {\bibfnamefont {L.~W.}\ \bibnamefont
  {Harriger}}, \bibinfo {author} {\bibfnamefont {H.~Q.}\ \bibnamefont {Luo}},
  \bibinfo {author} {\bibfnamefont {M.~S.}\ \bibnamefont {Liu}}, \bibinfo
  {author} {\bibfnamefont {C.}~\bibnamefont {Frost}}, \bibinfo {author}
  {\bibfnamefont {J.~P.}\ \bibnamefont {Hu}}, \bibinfo {author} {\bibfnamefont
  {M.~R.}\ \bibnamefont {Norman}}, \ and\ \bibinfo {author} {\bibfnamefont
  {P.}~\bibnamefont {Dai}},\ }\href@noop {} {\bibfield  {journal} {\bibinfo
  {journal} {Phys. Rev. B}\ }\textbf {\bibinfo {volume} {84}},\ \bibinfo
  {pages} {054544} (\bibinfo {year} {2011})}\BibitemShut {NoStop}%
\bibitem [{\citenamefont {Arham}\ \emph {et~al.}()\citenamefont {Arham},
  \citenamefont {Hunt}, \citenamefont {Park}, \citenamefont {Gillett},
  \citenamefont {Das}, \citenamefont {Sebastian}, \citenamefont {Xu},
  \citenamefont {Wen}, \citenamefont {Lin}, \citenamefont {Li}, \citenamefont
  {Gu}, \citenamefont {Thaler}, \citenamefont {Budko}, \citenamefont
  {Canfield},\ and\ \citenamefont {Greene}}]{Arham2011}%
  \BibitemOpen
  \bibfield  {author} {\bibinfo {author} {\bibfnamefont {H.~Z.}\ \bibnamefont
  {Arham}}, \bibinfo {author} {\bibfnamefont {C.~R.}\ \bibnamefont {Hunt}},
  \bibinfo {author} {\bibfnamefont {W.~K.}\ \bibnamefont {Park}}, \bibinfo
  {author} {\bibfnamefont {J.}~\bibnamefont {Gillett}}, \bibinfo {author}
  {\bibfnamefont {S.~D.}\ \bibnamefont {Das}}, \bibinfo {author} {\bibfnamefont
  {S.~E.}\ \bibnamefont {Sebastian}}, \bibinfo {author} {\bibfnamefont {Z.~J.}\
  \bibnamefont {Xu}}, \bibinfo {author} {\bibfnamefont {J.~S.}\ \bibnamefont
  {Wen}}, \bibinfo {author} {\bibfnamefont {Z.~W.}\ \bibnamefont {Lin}},
  \bibinfo {author} {\bibfnamefont {Q.}~\bibnamefont {Li}}, \bibinfo {author}
  {\bibfnamefont {G.}~\bibnamefont {Gu}}, \bibinfo {author} {\bibfnamefont
  {A.}~\bibnamefont {Thaler}}, \bibinfo {author} {\bibfnamefont {S.~L.}\
  \bibnamefont {Budko}}, \bibinfo {author} {\bibfnamefont {P.~C.}\ \bibnamefont
  {Canfield}}, \ and\ \bibinfo {author} {\bibfnamefont {L.~H.}\ \bibnamefont
  {Greene}},\ }\href@noop {} {\ }\Eprint {http://arxiv.org/abs/arXiv:1108.2749}
  {arXiv:1108.2749} \BibitemShut {NoStop}%
\bibitem [{\citenamefont {Song}\ \emph {et~al.}(2011)\citenamefont {Song},
  \citenamefont {Wang}, \citenamefont {Cheng}, \citenamefont {Jiang},
  \citenamefont {Li}, \citenamefont {Zhang}, \citenamefont {Li}, \citenamefont
  {He}, \citenamefont {Wang}, \citenamefont {Jia}, \citenamefont {Hung},
  \citenamefont {Wu}, \citenamefont {Ma}, \citenamefont {Chen},\ and\
  \citenamefont {Xue}}]{Song2011}%
  \BibitemOpen
  \bibfield  {author} {\bibinfo {author} {\bibfnamefont {C.-L.}\ \bibnamefont
  {Song}}, \bibinfo {author} {\bibfnamefont {Y.-L.}\ \bibnamefont {Wang}},
  \bibinfo {author} {\bibfnamefont {P.}~\bibnamefont {Cheng}}, \bibinfo
  {author} {\bibfnamefont {Y.-P.}\ \bibnamefont {Jiang}}, \bibinfo {author}
  {\bibfnamefont {W.}~\bibnamefont {Li}}, \bibinfo {author} {\bibfnamefont
  {T.}~\bibnamefont {Zhang}}, \bibinfo {author} {\bibfnamefont
  {Z.}~\bibnamefont {Li}}, \bibinfo {author} {\bibfnamefont {K.}~\bibnamefont
  {He}}, \bibinfo {author} {\bibfnamefont {L.}~\bibnamefont {Wang}}, \bibinfo
  {author} {\bibfnamefont {J.-F.}\ \bibnamefont {Jia}}, \bibinfo {author}
  {\bibfnamefont {H.-H.}\ \bibnamefont {Hung}}, \bibinfo {author}
  {\bibfnamefont {C.}~\bibnamefont {Wu}}, \bibinfo {author} {\bibfnamefont
  {X.}~\bibnamefont {Ma}}, \bibinfo {author} {\bibfnamefont {X.}~\bibnamefont
  {Chen}}, \ and\ \bibinfo {author} {\bibfnamefont {Q.-K.}\ \bibnamefont
  {Xue}},\ }\href@noop {} {\bibfield  {journal} {\bibinfo  {journal} {Science}\
  }\textbf {\bibinfo {volume} {332}},\ \bibinfo {pages} {1410} (\bibinfo {year}
  {2011})}\BibitemShut {NoStop}%
\bibitem [{\citenamefont {Si}\ and\ \citenamefont {Abrahams}(2008)}]{Si2008}%
  \BibitemOpen
  \bibfield  {author} {\bibinfo {author} {\bibfnamefont {Q.}~\bibnamefont
  {Si}}\ and\ \bibinfo {author} {\bibfnamefont {E.}~\bibnamefont {Abrahams}},\
  }\href@noop {} {\bibfield  {journal} {\bibinfo  {journal} {Phys. Rev. Lett.}\
  }\textbf {\bibinfo {volume} {101}},\ \bibinfo {pages} {076401} (\bibinfo
  {year} {2008})}\BibitemShut {NoStop}%
\bibitem [{\citenamefont {Xu}\ \emph {et~al.}(2008)\citenamefont {Xu},
  \citenamefont {M\"{u}ller},\ and\ \citenamefont {Sachdev}}]{Xu2008}%
  \BibitemOpen
  \bibfield  {author} {\bibinfo {author} {\bibfnamefont {C.}~\bibnamefont
  {Xu}}, \bibinfo {author} {\bibfnamefont {M.}~\bibnamefont {M\"{u}ller}}, \
  and\ \bibinfo {author} {\bibfnamefont {S.}~\bibnamefont {Sachdev}},\
  }\href@noop {} {\bibfield  {journal} {\bibinfo  {journal} {Phys. Rev. B}\
  }\textbf {\bibinfo {volume} {78}},\ \bibinfo {eid} {020501} (\bibinfo {year}
  {2008})}\BibitemShut {NoStop}%
\bibitem [{\citenamefont {Fang}\ \emph {et~al.}(2008)\citenamefont {Fang},
  \citenamefont {Yao}, \citenamefont {Tsai}, \citenamefont {Hu},\ and\
  \citenamefont {Kivelson}}]{Fang2008}%
  \BibitemOpen
  \bibfield  {author} {\bibinfo {author} {\bibfnamefont {C.}~\bibnamefont
  {Fang}}, \bibinfo {author} {\bibfnamefont {H.}~\bibnamefont {Yao}}, \bibinfo
  {author} {\bibfnamefont {W.-F.}\ \bibnamefont {Tsai}}, \bibinfo {author}
  {\bibfnamefont {J.}~\bibnamefont {Hu}}, \ and\ \bibinfo {author}
  {\bibfnamefont {S.~A.}\ \bibnamefont {Kivelson}},\ }\href@noop {} {\bibfield
  {journal} {\bibinfo  {journal} {Phys. Rev. B}\ }\textbf {\bibinfo {volume}
  {77}},\ \bibinfo {eid} {224509} (\bibinfo {year} {2008})}\BibitemShut
  {NoStop}%
\bibitem [{\citenamefont {Chandra}\ \emph {et~al.}(1990)\citenamefont
  {Chandra}, \citenamefont {Coleman},\ and\ \citenamefont
  {Larkin}}]{Chandra1990a}%
  \BibitemOpen
  \bibfield  {author} {\bibinfo {author} {\bibfnamefont {P.}~\bibnamefont
  {Chandra}}, \bibinfo {author} {\bibfnamefont {P.}~\bibnamefont {Coleman}}, \
  and\ \bibinfo {author} {\bibfnamefont {A.~I.}\ \bibnamefont {Larkin}},\
  }\href@noop {} {\bibfield  {journal} {\bibinfo  {journal} {Phys. Rev. Lett.}\
  }\textbf {\bibinfo {volume} {64}},\ \bibinfo {pages} {88} (\bibinfo {year}
  {1990})}\BibitemShut {NoStop}%
\bibitem [{\citenamefont {Zhai}\ \emph {et~al.}(2009)\citenamefont {Zhai},
  \citenamefont {Wang},\ and\ \citenamefont {Lee}}]{Zhai2009}%
  \BibitemOpen
  \bibfield  {author} {\bibinfo {author} {\bibfnamefont {H.}~\bibnamefont
  {Zhai}}, \bibinfo {author} {\bibfnamefont {F.}~\bibnamefont {Wang}}, \ and\
  \bibinfo {author} {\bibfnamefont {D.-H.}\ \bibnamefont {Lee}},\ }\href@noop
  {} {\bibfield  {journal} {\bibinfo  {journal} {Phys. Rev. B}\ }\textbf
  {\bibinfo {volume} {80}},\ \bibinfo {pages} {064517} (\bibinfo {year}
  {2009})}\BibitemShut {NoStop}%
\bibitem [{\citenamefont {Kang}\ and\ \citenamefont {Te\ifmmode \check{s}\else
  \v{s}\fi{}anovi\ifmmode~\acute{c}\else \'{c}\fi{}}(2011)}]{Kang2011}%
  \BibitemOpen
  \bibfield  {author} {\bibinfo {author} {\bibfnamefont {J.}~\bibnamefont
  {Kang}}\ and\ \bibinfo {author} {\bibfnamefont {Z.}~\bibnamefont {Te\ifmmode
  \check{s}\else \v{s}\fi{}anovi\ifmmode~\acute{c}\else \'{c}\fi{}}},\
  }\href@noop {} {\bibfield  {journal} {\bibinfo  {journal} {Phys. Rev. B}\
  }\textbf {\bibinfo {volume} {83}},\ \bibinfo {pages} {020505} (\bibinfo
  {year} {2011})}\BibitemShut {NoStop}%
\bibitem [{\citenamefont {Fernandes}\ \emph
  {et~al.}({\natexlab{a}})\citenamefont {Fernandes}, \citenamefont {Chubukov},
  \citenamefont {Knolle}, \citenamefont {Eremin},\ and\ \citenamefont
  {Schmalian}}]{Fernandes2011a}%
  \BibitemOpen
  \bibfield  {author} {\bibinfo {author} {\bibfnamefont {R.~M.}\ \bibnamefont
  {Fernandes}}, \bibinfo {author} {\bibfnamefont {A.~V.}\ \bibnamefont
  {Chubukov}}, \bibinfo {author} {\bibfnamefont {J.}~\bibnamefont {Knolle}},
  \bibinfo {author} {\bibfnamefont {I.}~\bibnamefont {Eremin}}, \ and\ \bibinfo
  {author} {\bibfnamefont {J.}~\bibnamefont {Schmalian}},\ }\href@noop {} {\ }
  \Eprint {http://arxiv.org/abs/arXiv:1110.1893}
  {arXiv:1110.1893} \BibitemShut {NoStop}%
\bibitem [{\citenamefont {Kivelson}\ \emph {et~al.}(1998)\citenamefont
  {Kivelson}, \citenamefont {Fradkin},\ and\ \citenamefont
  {Emery}}]{Kivelson1998}%
  \BibitemOpen
  \bibfield  {author} {\bibinfo {author} {\bibfnamefont {S.~A.}\ \bibnamefont
  {Kivelson}}, \bibinfo {author} {\bibfnamefont {E.}~\bibnamefont {Fradkin}}, \
  and\ \bibinfo {author} {\bibfnamefont {V.~J.}\ \bibnamefont {Emery}},\
  }\href@noop {} {\bibfield  {journal} {\bibinfo  {journal} {Nature}\ }\textbf
  {\bibinfo {volume} {393}},\ \bibinfo {pages} {550} (\bibinfo {year}
  {1998})}\BibitemShut {NoStop}%
\bibitem [{\citenamefont {Zeljkovic}\ \emph {et~al.}()\citenamefont
  {Zeljkovic}, \citenamefont {Main}, \citenamefont {Williams}, \citenamefont
  {Boyer}, \citenamefont {Chatterjee}, \citenamefont {Wise}, \citenamefont
  {Kondo}, \citenamefont {Takeuchi}, \citenamefont {Ikuta}, \citenamefont {Gu},
  \citenamefont {Hudson},\ and\ \citenamefont {Hoffman}}]{Zeljkovic2011}%
  \BibitemOpen
  \bibfield  {author} {\bibinfo {author} {\bibfnamefont {I.}~\bibnamefont
  {Zeljkovic}}, \bibinfo {author} {\bibfnamefont {E.~J.}\ \bibnamefont {Main}},
  \bibinfo {author} {\bibfnamefont {T.~L.}\ \bibnamefont {Williams}}, \bibinfo
  {author} {\bibfnamefont {M.~C.}\ \bibnamefont {Boyer}}, \bibinfo {author}
  {\bibfnamefont {K.}~\bibnamefont {Chatterjee}}, \bibinfo {author}
  {\bibfnamefont {W.~D.}\ \bibnamefont {Wise}}, \bibinfo {author}
  {\bibfnamefont {T.}~\bibnamefont {Kondo}}, \bibinfo {author} {\bibfnamefont
  {T.}~\bibnamefont {Takeuchi}}, \bibinfo {author} {\bibfnamefont
  {H.}~\bibnamefont {Ikuta}}, \bibinfo {author} {\bibfnamefont {G.~D.}\
  \bibnamefont {Gu}}, \bibinfo {author} {\bibfnamefont {E.~W.}\ \bibnamefont
  {Hudson}}, \ and\ \bibinfo {author} {\bibfnamefont {J.~E.}\ \bibnamefont
  {Hoffman}},\ }\href@noop {} {\ }\Eprint
  {http://arxiv.org/abs/arXiv:1104.4342} {arXiv:1104.4342} \BibitemShut
  {NoStop}%
\bibitem [{\citenamefont {Chen}\ \emph {et~al.}(2010)\citenamefont {Chen},
  \citenamefont {Maciejko}, \citenamefont {Sorini}, \citenamefont {Moritz},
  \citenamefont {Singh},\ and\ \citenamefont {Devereaux}}]{Chen2010}%
  \BibitemOpen
  \bibfield  {author} {\bibinfo {author} {\bibfnamefont {C.-C.}\ \bibnamefont
  {Chen}}, \bibinfo {author} {\bibfnamefont {J.}~\bibnamefont {Maciejko}},
  \bibinfo {author} {\bibfnamefont {A.~P.}\ \bibnamefont {Sorini}}, \bibinfo
  {author} {\bibfnamefont {B.}~\bibnamefont {Moritz}}, \bibinfo {author}
  {\bibfnamefont {R.~R.~P.}\ \bibnamefont {Singh}}, \ and\ \bibinfo {author}
  {\bibfnamefont {T.~P.}\ \bibnamefont {Devereaux}},\ }\href@noop {} {\bibfield
   {journal} {\bibinfo  {journal} {Phys. Rev. B}\ }\textbf {\bibinfo {volume}
  {82}},\ \bibinfo {pages} {100504} (\bibinfo {year} {2010})}\BibitemShut
  {NoStop}%
\bibitem [{\citenamefont {Lee}\ and\ \citenamefont {Wen}(2008)}]{Lee2008}%
  \BibitemOpen
  \bibfield  {author} {\bibinfo {author} {\bibfnamefont {P.~A.}\ \bibnamefont
  {Lee}}\ and\ \bibinfo {author} {\bibfnamefont {X.-G.}\ \bibnamefont {Wen}},\
  }\href@noop {} {\bibfield  {journal} {\bibinfo  {journal} {Phys. Rev. B}\
  }\textbf {\bibinfo {volume} {78}},\ \bibinfo {eid} {144517} (\bibinfo {year}
  {2008})}\BibitemShut {NoStop}%
\bibitem [{\citenamefont {Raghu}\ \emph {et~al.}(2008)\citenamefont {Raghu},
  \citenamefont {Qi}, \citenamefont {Liu}, \citenamefont {Scalapino},\ and\
  \citenamefont {Zhang}}]{Raghu2008}%
  \BibitemOpen
  \bibfield  {author} {\bibinfo {author} {\bibfnamefont {S.}~\bibnamefont
  {Raghu}}, \bibinfo {author} {\bibfnamefont {X.-L.}\ \bibnamefont {Qi}},
  \bibinfo {author} {\bibfnamefont {C.-X.}\ \bibnamefont {Liu}}, \bibinfo
  {author} {\bibfnamefont {D.~J.}\ \bibnamefont {Scalapino}}, \ and\ \bibinfo
  {author} {\bibfnamefont {S.-C.}\ \bibnamefont {Zhang}},\ }\href@noop {}
  {\bibfield  {journal} {\bibinfo  {journal} {Phys. Rev. B}\ }\textbf {\bibinfo
  {volume} {77}},\ \bibinfo {eid} {220503} (\bibinfo {year}
  {2008})}\BibitemShut {NoStop}%
\bibitem [{\citenamefont {Ran}\ \emph {et~al.}(2009)\citenamefont {Ran},
  \citenamefont {Wang}, \citenamefont {Zhai}, \citenamefont {Vishwanath},\ and\
  \citenamefont {Lee}}]{Ran2009}%
  \BibitemOpen
  \bibfield  {author} {\bibinfo {author} {\bibfnamefont {Y.}~\bibnamefont
  {Ran}}, \bibinfo {author} {\bibfnamefont {F.}~\bibnamefont {Wang}}, \bibinfo
  {author} {\bibfnamefont {H.}~\bibnamefont {Zhai}}, \bibinfo {author}
  {\bibfnamefont {A.}~\bibnamefont {Vishwanath}}, \ and\ \bibinfo {author}
  {\bibfnamefont {D.-H.}\ \bibnamefont {Lee}},\ }\href@noop {} {\bibfield
  {journal} {\bibinfo  {journal} {Phys. Rev. B}\ }\textbf {\bibinfo {volume}
  {79}},\ \bibinfo {pages} {014505} (\bibinfo {year} {2009})}\BibitemShut
  {NoStop}%
\bibitem [{\citenamefont {Moreo}\ \emph {et~al.}(2009)\citenamefont {Moreo},
  \citenamefont {Daghofer}, \citenamefont {Riera},\ and\ \citenamefont
  {Dagotto}}]{Moreo2009}%
  \BibitemOpen
  \bibfield  {author} {\bibinfo {author} {\bibfnamefont {A.}~\bibnamefont
  {Moreo}}, \bibinfo {author} {\bibfnamefont {M.}~\bibnamefont {Daghofer}},
  \bibinfo {author} {\bibfnamefont {J.~A.}\ \bibnamefont {Riera}}, \ and\
  \bibinfo {author} {\bibfnamefont {E.}~\bibnamefont {Dagotto}},\ }\href@noop
  {} {\bibfield  {journal} {\bibinfo  {journal} {Phys. Rev. B}\ }\textbf
  {\bibinfo {volume} {79}},\ \bibinfo {eid} {134502} (\bibinfo {year}
  {2009})}\BibitemShut {NoStop}%
\bibitem [{\citenamefont {Daghofer}\ \emph
  {et~al.}(2010{\natexlab{a}})\citenamefont {Daghofer}, \citenamefont
  {Nicholson}, \citenamefont {Moreo},\ and\ \citenamefont
  {Dagotto}}]{Daghofer2010}%
  \BibitemOpen
  \bibfield  {author} {\bibinfo {author} {\bibfnamefont {M.}~\bibnamefont
  {Daghofer}}, \bibinfo {author} {\bibfnamefont {A.}~\bibnamefont {Nicholson}},
  \bibinfo {author} {\bibfnamefont {A.}~\bibnamefont {Moreo}}, \ and\ \bibinfo
  {author} {\bibfnamefont {E.}~\bibnamefont {Dagotto}},\ }\href@noop {}
  {\bibfield  {journal} {\bibinfo  {journal} {Phys. Rev. B}\ }\textbf {\bibinfo
  {volume} {81}},\ \bibinfo {pages} {014511} (\bibinfo {year}
  {2010}{\natexlab{a}})}\BibitemShut {NoStop}%
\bibitem [{\citenamefont {Yu}\ \emph {et~al.}(2009)\citenamefont {Yu},
  \citenamefont {Trinh}, \citenamefont {Moreo}, \citenamefont {Daghofer},
  \citenamefont {Riera}, \citenamefont {Haas},\ and\ \citenamefont
  {Dagotto}}]{Yu2009}%
  \BibitemOpen
  \bibfield  {author} {\bibinfo {author} {\bibfnamefont {R.}~\bibnamefont
  {Yu}}, \bibinfo {author} {\bibfnamefont {K.~T.}\ \bibnamefont {Trinh}},
  \bibinfo {author} {\bibfnamefont {A.}~\bibnamefont {Moreo}}, \bibinfo
  {author} {\bibfnamefont {M.}~\bibnamefont {Daghofer}}, \bibinfo {author}
  {\bibfnamefont {J.~A.}\ \bibnamefont {Riera}}, \bibinfo {author}
  {\bibfnamefont {S.}~\bibnamefont {Haas}}, \ and\ \bibinfo {author}
  {\bibfnamefont {E.}~\bibnamefont {Dagotto}},\ }\href@noop {} {\bibfield
  {journal} {\bibinfo  {journal} {Phys. Rev. B}\ }\textbf {\bibinfo {volume}
  {79}},\ \bibinfo {pages} {104510} (\bibinfo {year} {2009})}\BibitemShut
  {NoStop}%
\bibitem [{\citenamefont {Kuroki}\ \emph {et~al.}(2008)\citenamefont {Kuroki},
  \citenamefont {Onari}, \citenamefont {Arita}, \citenamefont {Usui},
  \citenamefont {Tanaka}, \citenamefont {Kontani},\ and\ \citenamefont
  {Aoki}}]{Kuroki2008}%
  \BibitemOpen
  \bibfield  {author} {\bibinfo {author} {\bibfnamefont {K.}~\bibnamefont
  {Kuroki}}, \bibinfo {author} {\bibfnamefont {S.}~\bibnamefont {Onari}},
  \bibinfo {author} {\bibfnamefont {R.}~\bibnamefont {Arita}}, \bibinfo
  {author} {\bibfnamefont {H.}~\bibnamefont {Usui}}, \bibinfo {author}
  {\bibfnamefont {Y.}~\bibnamefont {Tanaka}}, \bibinfo {author} {\bibfnamefont
  {H.}~\bibnamefont {Kontani}}, \ and\ \bibinfo {author} {\bibfnamefont
  {H.}~\bibnamefont {Aoki}},\ }\href@noop {} {\bibfield  {journal} {\bibinfo
  {journal} {Phys. Rev. Lett.}\ }\textbf {\bibinfo {volume} {101}},\ \bibinfo
  {pages} {087004} (\bibinfo {year} {2008})}\BibitemShut {NoStop}%
\bibitem [{\citenamefont {Graser}\ \emph {et~al.}(2009)\citenamefont {Graser},
  \citenamefont {Maier}, \citenamefont {Hirschfeld},\ and\ \citenamefont
  {Scalapino}}]{Graser2009}%
  \BibitemOpen
  \bibfield  {author} {\bibinfo {author} {\bibfnamefont {S.}~\bibnamefont
  {Graser}}, \bibinfo {author} {\bibfnamefont {T.~A.}\ \bibnamefont {Maier}},
  \bibinfo {author} {\bibfnamefont {P.~J.}\ \bibnamefont {Hirschfeld}}, \ and\
  \bibinfo {author} {\bibfnamefont {D.~J.}\ \bibnamefont {Scalapino}},\
  }\href@noop {} {\bibfield  {journal} {\bibinfo  {journal} {New Journal of
  Physics}\ }\textbf {\bibinfo {volume} {11}},\ \bibinfo {pages} {025016}
  (\bibinfo {year} {2009})}\BibitemShut {NoStop}%
\bibitem [{\citenamefont {Calder\'on}\ \emph {et~al.}(2009)\citenamefont
  {Calder\'on}, \citenamefont {Valenzuela},\ and\ \citenamefont
  {Bascones}}]{Calderon2009}%
  \BibitemOpen
  \bibfield  {author} {\bibinfo {author} {\bibfnamefont {M.~J.}\ \bibnamefont
  {Calder\'on}}, \bibinfo {author} {\bibfnamefont {B.}~\bibnamefont
  {Valenzuela}}, \ and\ \bibinfo {author} {\bibfnamefont {E.}~\bibnamefont
  {Bascones}},\ }\href@noop {} {\bibfield  {journal} {\bibinfo  {journal}
  {Phys. Rev. B}\ }\textbf {\bibinfo {volume} {80}},\ \bibinfo {pages} {094531}
  (\bibinfo {year} {2009})}\BibitemShut {NoStop}%
\bibitem [{\citenamefont {Eschrig}\ and\ \citenamefont
  {Koepernik}(2009)}]{Eschrig2009}%
  \BibitemOpen
  \bibfield  {author} {\bibinfo {author} {\bibfnamefont {H.}~\bibnamefont
  {Eschrig}}\ and\ \bibinfo {author} {\bibfnamefont {K.}~\bibnamefont
  {Koepernik}},\ }\href@noop {} {\bibfield  {journal} {\bibinfo  {journal}
  {Phys. Rev. B}\ }\textbf {\bibinfo {volume} {80}},\ \bibinfo {pages} {104503}
  (\bibinfo {year} {2009})}\BibitemShut {NoStop}%
\bibitem [{\citenamefont {Graser}\ \emph {et~al.}(2010)\citenamefont {Graser},
  \citenamefont {Kemper}, \citenamefont {Maier}, \citenamefont {Cheng},
  \citenamefont {Hirschfeld},\ and\ \citenamefont {Scalapino}}]{Graser2010}%
  \BibitemOpen
  \bibfield  {author} {\bibinfo {author} {\bibfnamefont {S.}~\bibnamefont
  {Graser}}, \bibinfo {author} {\bibfnamefont {A.~F.}\ \bibnamefont {Kemper}},
  \bibinfo {author} {\bibfnamefont {T.~A.}\ \bibnamefont {Maier}}, \bibinfo
  {author} {\bibfnamefont {H.-P.}\ \bibnamefont {Cheng}}, \bibinfo {author}
  {\bibfnamefont {P.~J.}\ \bibnamefont {Hirschfeld}}, \ and\ \bibinfo {author}
  {\bibfnamefont {D.~J.}\ \bibnamefont {Scalapino}},\ }\href@noop {} {\bibfield
   {journal} {\bibinfo  {journal} {Phys. Rev. B}\ }\textbf {\bibinfo {volume}
  {81}},\ \bibinfo {pages} {214503} (\bibinfo {year} {2010})}\BibitemShut
  {NoStop}%
\bibitem [{\citenamefont {Luo}\ \emph {et~al.}(2010)\citenamefont {Luo},
  \citenamefont {Martins}, \citenamefont {Yao}, \citenamefont {Daghofer},
  \citenamefont {Yu}, \citenamefont {Moreo},\ and\ \citenamefont
  {Dagotto}}]{Luo2010}%
  \BibitemOpen
  \bibfield  {author} {\bibinfo {author} {\bibfnamefont {Q.}~\bibnamefont
  {Luo}}, \bibinfo {author} {\bibfnamefont {G.}~\bibnamefont {Martins}},
  \bibinfo {author} {\bibfnamefont {D.-X.}\ \bibnamefont {Yao}}, \bibinfo
  {author} {\bibfnamefont {M.}~\bibnamefont {Daghofer}}, \bibinfo {author}
  {\bibfnamefont {R.}~\bibnamefont {Yu}}, \bibinfo {author} {\bibfnamefont
  {A.}~\bibnamefont {Moreo}}, \ and\ \bibinfo {author} {\bibfnamefont
  {E.}~\bibnamefont {Dagotto}},\ }\href@noop {} {\bibfield  {journal} {\bibinfo
   {journal} {Phys. Rev. B}\ }\textbf {\bibinfo {volume} {82}},\ \bibinfo
  {pages} {104508} (\bibinfo {year} {2010})}\BibitemShut {NoStop}%
\bibitem [{\citenamefont {Fernandes}\ \emph
  {et~al.}({\natexlab{b}})\citenamefont {Fernandes}, \citenamefont {Abrahams},\
  and\ \citenamefont {Schmalian}}]{Fernandes2011}%
  \BibitemOpen
  \bibfield  {author} {\bibinfo {author} {\bibfnamefont {R.~M.}\ \bibnamefont
  {Fernandes}}, \bibinfo {author} {\bibfnamefont {E.}~\bibnamefont {Abrahams}},
  \ and\ \bibinfo {author} {\bibfnamefont {J.}~\bibnamefont {Schmalian}},\
  }\href@noop {} {\ }\Eprint
  {http://arxiv.org/abs/arXiv:1105.3906} {arXiv:1105.3906} \BibitemShut
  {NoStop}%
\bibitem [{\citenamefont {Kaneshita}\ \emph {et~al.}(2009)\citenamefont
  {Kaneshita}, \citenamefont {Morinari},\ and\ \citenamefont
  {Tohyama}}]{Kaneshita2009}%
  \BibitemOpen
  \bibfield  {author} {\bibinfo {author} {\bibfnamefont {E.}~\bibnamefont
  {Kaneshita}}, \bibinfo {author} {\bibfnamefont {T.}~\bibnamefont {Morinari}},
  \ and\ \bibinfo {author} {\bibfnamefont {T.}~\bibnamefont {Tohyama}},\
  }\href@noop {} {\bibfield  {journal} {\bibinfo  {journal} {Phys. Rev. Lett.}\
  }\textbf {\bibinfo {volume} {103}},\ \bibinfo {pages} {247202} (\bibinfo
  {year} {2009})}\BibitemShut {NoStop}%
\bibitem [{\citenamefont {Sugimoto}\ \emph {et~al.}(2011)\citenamefont
  {Sugimoto}, \citenamefont {Kaneshita},\ and\ \citenamefont
  {Tohyama}}]{Sugimoto2011}%
  \BibitemOpen
  \bibfield  {author} {\bibinfo {author} {\bibfnamefont {K.}~\bibnamefont
  {Sugimoto}}, \bibinfo {author} {\bibfnamefont {E.}~\bibnamefont {Kaneshita}},
  \ and\ \bibinfo {author} {\bibfnamefont {T.}~\bibnamefont {Tohyama}},\
  }\href@noop {} {\bibfield  {journal} {\bibinfo  {journal} {Journal of the
  Physical Society of Japan}\ }\textbf {\bibinfo {volume} {80}},\ \bibinfo
  {pages} {033706} (\bibinfo {year} {2011})}\BibitemShut {NoStop}%
\bibitem [{\citenamefont {Bascones}\ \emph {et~al.}(2010)\citenamefont
  {Bascones}, \citenamefont {Calder\'on},\ and\ \citenamefont
  {Valenzuela}}]{Bascones2010}%
  \BibitemOpen
  \bibfield  {author} {\bibinfo {author} {\bibfnamefont {E.}~\bibnamefont
  {Bascones}}, \bibinfo {author} {\bibfnamefont {M.~J.}\ \bibnamefont
  {Calder\'on}}, \ and\ \bibinfo {author} {\bibfnamefont {B.}~\bibnamefont
  {Valenzuela}},\ }\href@noop {} {\bibfield  {journal} {\bibinfo  {journal}
  {Phys. Rev. Lett.}\ }\textbf {\bibinfo {volume} {104}},\ \bibinfo {pages}
  {227201} (\bibinfo {year} {2010})}\BibitemShut {NoStop}%
\bibitem [{\citenamefont {Nevidomskyy}()}]{Nevidomskyy2011}%
  \BibitemOpen
  \bibfield  {author} {\bibinfo {author} {\bibfnamefont {A.~H.}\ \bibnamefont
  {Nevidomskyy}},\ }\href@noop {} {\ }\Eprint
  {http://arxiv.org/abs/arXiv:1104.1747} {arXiv:1104.1747} \BibitemShut
  {NoStop}%
\bibitem [{\citenamefont {Richard}\ \emph {et~al.}(2010)\citenamefont
  {Richard}, \citenamefont {Nakayama}, \citenamefont {Sato}, \citenamefont
  {Neupane}, \citenamefont {Xu}, \citenamefont {Bowen}, \citenamefont {Chen},
  \citenamefont {Luo}, \citenamefont {Wang}, \citenamefont {Dai}, \citenamefont
  {Fang}, \citenamefont {Ding},\ and\ \citenamefont {Takahashi}}]{Richard2010}%
  \BibitemOpen
  \bibfield  {author} {\bibinfo {author} {\bibfnamefont {P.}~\bibnamefont
  {Richard}}, \bibinfo {author} {\bibfnamefont {K.}~\bibnamefont {Nakayama}},
  \bibinfo {author} {\bibfnamefont {T.}~\bibnamefont {Sato}}, \bibinfo {author}
  {\bibfnamefont {M.}~\bibnamefont {Neupane}}, \bibinfo {author} {\bibfnamefont
  {Y.-M.}\ \bibnamefont {Xu}}, \bibinfo {author} {\bibfnamefont {J.~H.}\
  \bibnamefont {Bowen}}, \bibinfo {author} {\bibfnamefont {G.~F.}\ \bibnamefont
  {Chen}}, \bibinfo {author} {\bibfnamefont {J.~L.}\ \bibnamefont {Luo}},
  \bibinfo {author} {\bibfnamefont {N.~L.}\ \bibnamefont {Wang}}, \bibinfo
  {author} {\bibfnamefont {X.}~\bibnamefont {Dai}}, \bibinfo {author}
  {\bibfnamefont {Z.}~\bibnamefont {Fang}}, \bibinfo {author} {\bibfnamefont
  {H.}~\bibnamefont {Ding}}, \ and\ \bibinfo {author} {\bibfnamefont
  {T.}~\bibnamefont {Takahashi}},\ }\href@noop {} {\bibfield  {journal}
  {\bibinfo  {journal} {Phys. Rev. Lett.}\ }\textbf {\bibinfo {volume} {104}},\
  \bibinfo {pages} {137001} (\bibinfo {year} {2010})}\BibitemShut {NoStop}%
\bibitem [{\citenamefont {Yin}\ \emph {et~al.}(2011)\citenamefont {Yin},
  \citenamefont {Haule},\ and\ \citenamefont {Kotliar}}]{Yin2011}%
  \BibitemOpen
  \bibfield  {author} {\bibinfo {author} {\bibfnamefont {Z.~P.}\ \bibnamefont
  {Yin}}, \bibinfo {author} {\bibfnamefont {K.}~\bibnamefont {Haule}}, \ and\
  \bibinfo {author} {\bibfnamefont {G.}~\bibnamefont {Kotliar}},\ }\href@noop
  {} {\bibfield  {journal} {\bibinfo  {journal} {Nature Physics}\ }\textbf
  {\bibinfo {volume} {7}},\ \bibinfo {pages} {294} (\bibinfo {year}
  {2011})}\BibitemShut {NoStop}%
\bibitem [{\citenamefont {Daghofer}\ \emph
  {et~al.}(2010{\natexlab{b}})\citenamefont {Daghofer}, \citenamefont {Luo},
  \citenamefont {Yu}, \citenamefont {Yao}, \citenamefont {Moreo},\ and\
  \citenamefont {Dagotto}}]{Daghofer2010a}%
  \BibitemOpen
  \bibfield  {author} {\bibinfo {author} {\bibfnamefont {M.}~\bibnamefont
  {Daghofer}}, \bibinfo {author} {\bibfnamefont {Q.-L.}\ \bibnamefont {Luo}},
  \bibinfo {author} {\bibfnamefont {R.}~\bibnamefont {Yu}}, \bibinfo {author}
  {\bibfnamefont {D.~X.}\ \bibnamefont {Yao}}, \bibinfo {author} {\bibfnamefont
  {A.}~\bibnamefont {Moreo}}, \ and\ \bibinfo {author} {\bibfnamefont
  {E.}~\bibnamefont {Dagotto}},\ }\href@noop {} {\bibfield  {journal} {\bibinfo
   {journal} {Phys. Rev. B}\ }\textbf {\bibinfo {volume} {81}},\ \bibinfo
  {pages} {180514} (\bibinfo {year} {2010}{\natexlab{b}})}\BibitemShut
  {NoStop}%
\bibitem [{\citenamefont {Valenzuela}\ \emph {et~al.}(2010)\citenamefont
  {Valenzuela}, \citenamefont {Bascones},\ and\ \citenamefont
  {Calder\'on}}]{Valenzuela2010}%
  \BibitemOpen
  \bibfield  {author} {\bibinfo {author} {\bibfnamefont {B.}~\bibnamefont
  {Valenzuela}}, \bibinfo {author} {\bibfnamefont {E.}~\bibnamefont
  {Bascones}}, \ and\ \bibinfo {author} {\bibfnamefont {M.~J.}\ \bibnamefont
  {Calder\'on}},\ }\href@noop {} {\bibfield  {journal} {\bibinfo  {journal}
  {Phys. Rev. Lett.}\ }\textbf {\bibinfo {volume} {105}},\ \bibinfo {pages}
  {207202} (\bibinfo {year} {2010})}\BibitemShut {NoStop}%
\bibitem [{\citenamefont {Harrison}\ \emph {et~al.}(2007)\citenamefont
  {Harrison}, \citenamefont {McDonald},\ and\ \citenamefont
  {Singleton}}]{Harrison2007}%
  \BibitemOpen
  \bibfield  {author} {\bibinfo {author} {\bibfnamefont {N.}~\bibnamefont
  {Harrison}}, \bibinfo {author} {\bibfnamefont {R.~D.}\ \bibnamefont
  {McDonald}}, \ and\ \bibinfo {author} {\bibfnamefont {J.}~\bibnamefont
  {Singleton}},\ }\href@noop {} {\bibfield  {journal} {\bibinfo  {journal}
  {Phys. Rev. Lett.}\ }\textbf {\bibinfo {volume} {99}},\ \bibinfo {pages}
  {206406} (\bibinfo {year} {2007})}\BibitemShut {NoStop}%
\bibitem [{\citenamefont {Jesche}\ \emph {et~al.}(2010)\citenamefont {Jesche},
  \citenamefont {Krellner}, \citenamefont {de~Souza}, \citenamefont {Lang},\
  and\ \citenamefont {Geibel}}]{Jesche2010}%
  \BibitemOpen
  \bibfield  {author} {\bibinfo {author} {\bibfnamefont {A.}~\bibnamefont
  {Jesche}}, \bibinfo {author} {\bibfnamefont {C.}~\bibnamefont {Krellner}},
  \bibinfo {author} {\bibfnamefont {M.}~\bibnamefont {de~Souza}}, \bibinfo
  {author} {\bibfnamefont {M.}~\bibnamefont {Lang}}, \ and\ \bibinfo {author}
  {\bibfnamefont {C.}~\bibnamefont {Geibel}},\ }\href@noop {} {\bibfield
  {journal} {\bibinfo  {journal} {Phys. Rev. B}\ }\textbf {\bibinfo {volume}
  {81}},\ \bibinfo {pages} {134525} (\bibinfo {year} {2010})}\BibitemShut
  {NoStop}%
\bibitem [{\citenamefont {Liu}\ \emph {et~al.}(2011)\citenamefont {Liu},
  \citenamefont {Quan}, \citenamefont {Chen}, \citenamefont {Zou},\ and\
  \citenamefont {Lin}}]{Liu2011}%
  \BibitemOpen
  \bibfield  {author} {\bibinfo {author} {\bibfnamefont {D.-Y.}\ \bibnamefont
  {Liu}}, \bibinfo {author} {\bibfnamefont {Y.-M.}\ \bibnamefont {Quan}},
  \bibinfo {author} {\bibfnamefont {D.-M.}\ \bibnamefont {Chen}}, \bibinfo
  {author} {\bibfnamefont {L.-J.}\ \bibnamefont {Zou}}, \ and\ \bibinfo
  {author} {\bibfnamefont {H.-Q.}\ \bibnamefont {Lin}},\ }\href@noop {}
  {\bibfield  {journal} {\bibinfo  {journal} {Phys. Rev. B}\ }\textbf {\bibinfo
  {volume} {84}},\ \bibinfo {pages} {064435} (\bibinfo {year}
  {2011})}\BibitemShut {NoStop}%
\bibitem [{\citenamefont {Laad}\ and\ \citenamefont {Craco}(2011)}]{Laad2011}%
  \BibitemOpen
  \bibfield  {author} {\bibinfo {author} {\bibfnamefont {M.~S.}\ \bibnamefont
  {Laad}}\ and\ \bibinfo {author} {\bibfnamefont {L.}~\bibnamefont {Craco}},\
  }\href@noop {} {\bibfield  {journal} {\bibinfo  {journal} {Phys. Rev. B}\
  }\textbf {\bibinfo {volume} {84}},\ \bibinfo {pages} {054530} (\bibinfo
  {year} {2011})}\BibitemShut {NoStop}%
\end{thebibliography}%
\bibliographystyle{apsrev4-1}

\end{document}